\documentclass[twocolumn,prl,reprint,superscriptaddress]{revtex4}
\usepackage{amsmath,amssymb}
\usepackage{graphicx}
\usepackage{float}
\usepackage{subfigure}
\usepackage{verbatim}

\usepackage{amsfonts}
\usepackage{dcolumn}
\usepackage{bm}
\usepackage{array,color}
\usepackage[colorlinks,citecolor=blue]{hyperref}
\usepackage[english]{babel}

\begin{document}
\hyphenpenalty=5000
\tolerance=1000

\title{Demonstration of Einstein-Podolsky-Rosen Steering with Enhanced Subchannel Discrimination}

\renewcommand{\thefootnote}{\fnsymbol{footnote}}

\author{Kai Sun$^*$}
\affiliation{CAS Key Laboratory of Quantum Information, University of Science and Technology of China, Hefei 230026, People's Republic of China}
\affiliation{Synergetic Innovation Center of Quantum Information and Quantum Physics, University of Science and Technology of China, Hefei 230026, People's Republic of China}

\author{Xiang-Jun Ye\footnote{These two authors contributed equally to this work.}}
\affiliation{CAS Key Laboratory of Quantum Information, University of Science and Technology of China, Hefei 230026, People's Republic of China}
\affiliation{Synergetic Innovation Center of Quantum Information and Quantum Physics, University of Science and Technology of China, Hefei 230026, People's Republic of China}

\author{Ya Xiao}
\affiliation{CAS Key Laboratory of Quantum Information, University of Science and Technology of China, Hefei 230026, People's Republic of China}
\affiliation{Synergetic Innovation Center of Quantum Information and Quantum Physics, University of Science and Technology of China, Hefei 230026, People's Republic of China}

\author{Xiao-Ye Xu}
\affiliation{CAS Key Laboratory of Quantum Information, University of Science and Technology of China, Hefei 230026, People's Republic of China}
\affiliation{Synergetic Innovation Center of Quantum Information and Quantum Physics, University of Science and Technology of China, Hefei 230026, People's Republic of China}

\author{Yu-Chun Wu}
\affiliation{CAS Key Laboratory of Quantum Information, University of Science and Technology of China, Hefei 230026, People's Republic of China}
\affiliation{Synergetic Innovation Center of Quantum Information and Quantum Physics, University of Science and Technology of China, Hefei 230026, People's Republic of China}

\author{Jin-Shi Xu}
\email{jsxu@ustc.edu.cn}
\affiliation{CAS Key Laboratory of Quantum Information, University of Science and Technology of China, Hefei 230026, People's Republic of China}
\affiliation{Synergetic Innovation Center of Quantum Information and Quantum Physics, University of Science and Technology of China, Hefei 230026, People's Republic of China}

\author{Jing-Ling Chen}
\email{chenjl@nankai.edu.cn}
\affiliation{Theoretical Physics Division, Chern Institute of Mathematics, Nankai University, Tianjin, 300071, People's Republic of China}
\affiliation{Centre for Quantum Technologies, National University of Singapore, 3 Science Drive 2, Singapore, 117543}

\author{Chuan-Feng~Li}
\email{cfli@ustc.edu.cn}
\affiliation{CAS Key Laboratory of Quantum Information, University of Science and Technology of China, Hefei 230026, People's Republic of China}
\affiliation{Synergetic Innovation Center of Quantum Information and Quantum Physics, University of Science and Technology of China, Hefei 230026, People's Republic of China}

\author{Guang-Can Guo}
\affiliation{CAS Key Laboratory of Quantum Information, University of Science and Technology of China, Hefei 230026, People's Republic of China}
\affiliation{Synergetic Innovation Center of Quantum Information and Quantum Physics, University of Science and Technology of China, Hefei 230026, People's Republic of China}

\date{\today }

\begin{abstract}
{\bf Abstract}\ \ \ Einstein-Podolsky-Rosen (EPR) steering describes a quantum nonlocal phenomenon in which one party can nonlocally affect the other's state through local measurements. It reveals an additional concept of quantum nonlocality, which stands between quantum entanglement and Bell nonlocality. Recently, a quantum information task named as subchannel discrimination (SD) provides a necessary and sufficient characterization of EPR steering. The success probability of SD using steerable states is higher than using any unsteerable states, even when they are entangled. However, the detailed construction of such subchannels and the experimental realization of the corresponding task are still technologically challenging. In this work, we designed a feasible collection of subchannels for a quantum channel and experimentally demonstrated the corresponding SD task where the probabilities of correct discrimination are clearly enhanced by exploiting steerable states. Our results provide a concrete example to operationally demonstrate EPR steering and shine a new light on the potential application of EPR steering.\\

\end{abstract}

\maketitle

\section{Introduction}

\begin{figure*}[htbp]
  \centering
  \includegraphics[width=0.87\textwidth]{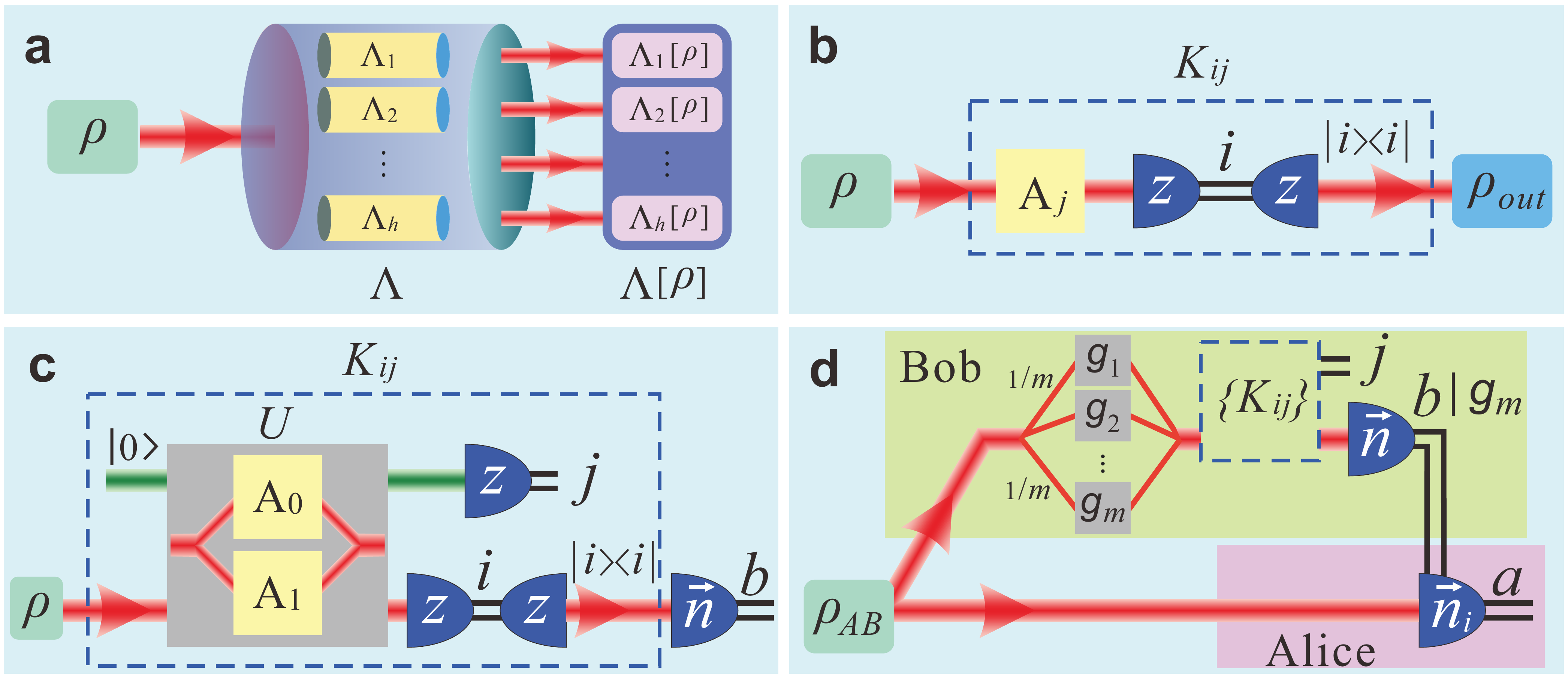}\\
  \caption{The process of subchannel discrimination (SD). {\bf a} The collection of subchannels $\{\Lambda_{h}\}_h$ composing a quantum channel $\Lambda$. {\bf b} The protocol for realizing the Kraus operators $K_{ij}$ using the entanglement-breaking channel (EBC) scenario. The state $\rho$ is measured along the $z$ direction with output result $i$ after passing through the intermediate subchannel $A_j$. A new state is prepared based on the value of $i$. {\bf c} The single-qubit protocol for SD in the case of two measurement settings. The unitary operation $U$ related to the intermediate subchannels $A_0$ and $A_1$ is demonstrated with the qubit state $\rho$ and an auxiliary qubit $|0\rangle$ which is finally measured along $z$ with output result $j$. After the EBC, the measurement along a direction $\vec{n}$ is performed on the signal qubit, and the result $b$ is obtained. {\bf d} The two-qubit protocol for SD with multiple measurement settings. One of the two qubits, in a state $\rho_{AB}$, is sent to Bob, and the other is sent to Alice. On Bob's side, the qubit passes through one of the unitary gates $g_m$ with probability $1/m$ before the evolution $K_{ij}$. Alice chooses one of the measurement directions $\vec{n}_i$ based on the result $b|g_m$ from Bob and obtains the result $a$.}\label{theory}
\end{figure*}

In the original discussion of Einstein-Podolsky-Rosen (EPR) paradox \cite{epr1935}, Schr\"{o}dinger \cite{schr1935,schr1936} described a quantum nonlocal phenomenon that Alice can steer Bob's state through her local measurements. Since then, great efforts have been made to understand quantum nonlocality. It was not until 2007, H. Wiseman, S. Jones and A. Doherty revisited Schr\"{o}dinger's discussion and formulated the concepts of quantum nonlocality as quantum entanglement, EPR steering and Bell nonlocality in terms of quantum information tasks \cite{wjd2007,jones2007}. It is now clear that all steerable states are entangled, but not all steerable states exhibit the Bell nonlocality \cite{wjd2007,jones2007}, which implies that EPR steering sits between quantum entanglement and Bell nonlocality. This hierarchy also holds for all possible positive operator valued measures \cite{quintino2015}. EPR steering has recently drawn plenty of attention \cite{caval2017}. For example, several theoretical studies including the verification of EPR steering based on steering inequalities \cite{criteria2009} and all-versus-nothing proof \cite{chenjl2013}, no-cloning of quantum steering \cite{chli2016}, temporal steering \cite{chenli2013,chenli2014,nori2016,chenli2016,chenli2017}, quantification of steerability \cite{paul2014,piani2015,costa2016} and one-way EPR steering \cite{joseph2014} have been reported as well as the corresponding experiments \cite{saunders2010,sun2014,norif2016,liu2017,wollmann2016,sun2016,xiao2017}. There are also other interesting steering experiments, such as the high-order steering \cite{panli2015} and loophole-free steering \cite{smith2012,bennet2012,wittmann2012}. Moreover, the parallel works based on the continuous variable systems \cite{reid1989,olsen2008,mdigley2010,vitus2012,he2013multi,he,he2015tele,kogias2015,seiji2015} have been reported.

Similar to the necessary and sufficient verification of quantum entanglement with a quantum information task named quantum channel discrimination \cite{piani2009} which refers to the task of distinguishing among different quantum operations \cite{acin2001,paris2001,seth2008}, EPR steering can be characterized necessarily and sufficiently based on a quantum task named subchannel discrimination (SD) \cite{piani2015}. As an extension of the quantum channel generally representing the physical transformation of information from an initial state to a final state in which the quantum operation is trace-preserving for all input states \cite{chuang2010}, a subchannel is a completely positive operator that does not increase the trace in the density matrix space \cite{piani2015}. A series of subchannels $\{\Lambda_h\}_h$, that constitute a channel $\Lambda$ satisfying $\Lambda=\sum_{h}{\Lambda_{h}}$, can be treated as a decomposition of the channel into its different evolutionary branches with the corresponding probability $\text{Tr}(\Lambda_{h}[\rho])$ for any state $\rho$, as shown in Fig. \ref{theory} {\bf a}. Here, $\Lambda_{h}[\rho]=K_{h}\rho K_{h}^\dag$, where the Kraus operators $K_{h}$ are the explicit matrix descriptions of $\Lambda_h$ and satisfy $\sum_{h} K_{h}^\dag K_{h}=\mathbb{I}$. The SD task allows one to distinguish in which subchannel the quantum evolution occurs, whereas this information is lost if the process is described simply in the framework of the quantum channel. Moreover, SD tasks might lead to the emergence of new quantum phenomena and applications in quantum information processing, such as the SD-based quantum key distribution \cite{cyril2012}.

Recently, it has been proven that for any bipartite state, we can verify it is steerable if there exists an SD task in which the successful discrimination probability is enhanced by this state compared with the case employing single-qubit states; otherwise, if no such SD tasks exist, it is unsteerable \cite{piani2015}. Also, such an SD task presents an operational method to characterize EPR steering. However, the detailed construction of such subchannels has not been investigated up to now. In this article, we design a feasible collection of concrete subchannels and experimentally demonstrate EPR steering with the corresponding SD task.

\section{Results}

{\it SD task for the two-setting case.} First, we would like to introduce the detailed SD task in the simplest case with two measurement settings. In this work, we consider a channel consisting of four subchannels $\Lambda_{ij}$ ($i,\,j=0$ or $1$), where the corresponding Kraus operators are denoted by $K_{ij}$. We exploit an entanglement-breaking channel (EBC) \cite{shor2003} to limit the bound established in the single-qubit protocol. The Kraus operators $K_{ij}$ are implemented with the EBC, as illustrated in Fig. \ref{theory} {\bf b}, where $A_j$ ($j=0$ or $1$) is regarded as the intermediate subchannel and satisfies $K_{i,j}=|i\rangle\langle i|\cdot A_j$ ($i,\,j=0$ or $1$) (see Methods). Since the information of $i$ is included in the output $\rho_{out}$, the SD task is transformed into the task of distinguishing $A_j$ based on $i$. To realize $\{A_j\}_j$, a unitary operation $U$ is performed on a quantum system consisting of a target qubit in the state $\rho$ and an auxiliary qubit initially in the state $|0\rangle$ \cite{stinespring1955}, as shown in Fig. \ref{theory} {\bf c}. In this work, the operation is represented as follows,
\begin{equation}\label{ugate}
U=\left(\begin{array}{cc}
A_0 & -A_1\\
A_1 & A_0\\
\end{array}\right).
\end{equation}
$A_j$ is determined according to the output $j$ measured along the $z$ direction on the auxiliary qubit. The SD task in single-qubit protocol is completed by guessing $j$ according to the output $b$ that is measured along a direction $\vec{n}$ on the target qubit. Since the target qubit only carries the classical information after the EBC, $\vec{n}$ is optimized to be $z$ to maximize the success probability $P^s_{\rho}$. With the input state $\rho$, the results of different strategies for guessing $j$ are denoted by $p_{\rho}^{c0},\,p_{\rho}^{c1}$ (guessing $j$ is the constant $0$ or $1$ regardless of $b$, respectively), $p_{\rho}^{00},\,p_{\rho}^{01}$ (guessing $j=b$ or $j=b\oplus 1$ where $\oplus$ represents addition modulo 2, respectively). The success probability is denoted as $P^s_{\rho}=\max{\{p_{\rho}^{c0},\,p_{\rho}^{c1},\,p_{\rho}^{00},\,p_{\rho}^{01}\}}$, and the upper-bound probability $P^s$ in the single-qubit case is obtained by optimizing the input state, which implies $P^s=\max_{\rho}{\{P^s_{\rho}\}}$.

We now consider the two-qubit Werner states $\rho_{AB}$ \cite{werner1989} with the form of,
\begin{equation}\label{rho1}
\rho_{AB}=\eta\,|\Phi\rangle\langle\Phi|+(1-\eta)\, \mathbb{I}/4,
\end{equation}
where $\eta\in[0,1]$, $|\Phi\rangle$ is the maximally entangled state, and $\mathbb{I}/4$ is the maximally mixed state. As illustrated in Fig. \ref{theory} {\bf d} in the two-setting case ($m=1$, and $g_1$ is identical), the task is that Alice guesses $j$ and announces to Bob based on $a$ which is obtained by measuring along $\vec{n}_i$ (chosen according to $b$). Since $b\in\{0,\,1\}$, there are two directions $\vec{n}_i$ along which Alice can choose to measure. In this work, we follow two rules to design the SD tasks, i.e., (i) the success probability of maximally entangled state is $100\%$; (ii) the success probability of maximally mixed state is $50\%$. Thus, the success probability of SD task $P_{\rho_{AB}}$ equals to $1/2+\eta/2$. In the linear EPR steering inequalities, $C_n^{LHS}$ denotes the bound established by the local hidden state model where $n$ is the number of measurement settings \cite{saunders2010}. In the case of $n=2$, $C^{LHS}_2=\eta_2^*$ where $\eta_2^*=1/\sqrt{2}$ is the visibility bound of the Werner states. When $\eta>\eta_2^*$, $\rho_{\rho_{AB}}$ is steerable. For the single-qubit protocol, by directly calculating, we find $P^s=1/2+C^{LHS}_2/2$ (see Section I of Appendix). Thus, if Bob finds $P_{\rho_{AB}}>P^s$, the steerability from Alice to Bob is observed.

{\it SD task for the multi-setting cases.} EPR steering from Alice to Bob relates to the number of settings measured by Alice \cite{wjd2007,evans2014}. For some predictably steerable states, steering fails because of the very limited number of measurement settings \cite{sun2016}. To capture as much information about the states as possible to demonstrate EPR steering, it is necessary for Alice to apply multiple measurement settings to approach the predictions of infinite measurement settings. In this work, we consider the regularly spaced directions which are given by the Platonic solids with the number of measurement settings $n$ corresponding to 2, 3, 4, 6 and 10 \cite{saunders2010}. Compared with the optimal measurement settings introduced in \cite{evans2014}, here, the two- and three-setting measurements are optimal and EPR steering can be affirmed necessarily and sufficiently. For other multi-setting cases, the optimal measurements don't correspond to the regularly spaced directions and are difficult to realize in experiment. Moreover, the diffence between the results in this work and the predictions of the optimal measurements is very small (see Section I of Appendix). To experimentally realize such a task, on Bob's side, the qubit equiprobably evolves through several unitary gates $g_m$ before the Kraus operators $K_{ij}$, as illustrated in Fig. \ref{theory} {\bf d}, and the details can be found in Section I of Appendix. The corresponding measurement setting based on Bob's result $b|g_m$, which denotes that $b$ is obtained under the gate operation $g_m$, is then implemented on Alice's side. In fact, for each $g_m$, the SD process can still be regarded within the framework of two measurement settings. In the single-qubit protocol with $n$ measurement settings, denoting $P^s_{m,\rho}$ for each $g_m$, the total success probability is obtained as $P^s_n=\max_{\rho}\{\sum_m 1/m\,P^s_{\rho,\,m}\}$. Similarly, for the two-qubit state $\rho_{AB}$, $P_{\rho_{AB},\,n}=\sum_m 1/m\,P_{\rho_{AB},\,m}$. Furthermore, similar with the two-setting case, we have $C^{LHS}_n=\eta_n^*$ and $P^s_n=1/2+C^{LHS}_n/2$ (see Section I of Appendix for details) when the multiple measurement settings are selected based on the Platonic solids \cite{saunders2010}. As a result, the constructed SD task provides an operational method to characterize the steerability of Werner states. If the success probability of SD is enhanced by using the two-qubit state $\rho_{AB}$, i.e., $P_{\rho_{AB}}>P^s$, then $\rho_{AB}$ is steerable from Alice to Bob regardless of the number of measurement settings; otherwise, under $n$ measurement settings performed by Alice, i.e., $P_{\rho_{AB},\,n}\leq P_n^s$, Alice fails to steer Bob's states.

Compared with the two-setting case in which there are four subchannels, the multi-setting cases can be regarded as multi-subchannel discrimination tasks where more subchannels consisting of gates $g_m$ and the corresponding Kraus operators $K_{ij}$ are required, and the reconstructed subchannels could be expressed as $K_{ijm}'=K_{ij}\cdot g_m$. Following the similar method designing subchannels for Werner states, we can also create the corresponding subchannels for other types of two-qubit states, like the Bell diagonal states (see Section I of Appendix).

\begin{figure}[!h]
\centering
\includegraphics[width=0.47\textwidth]{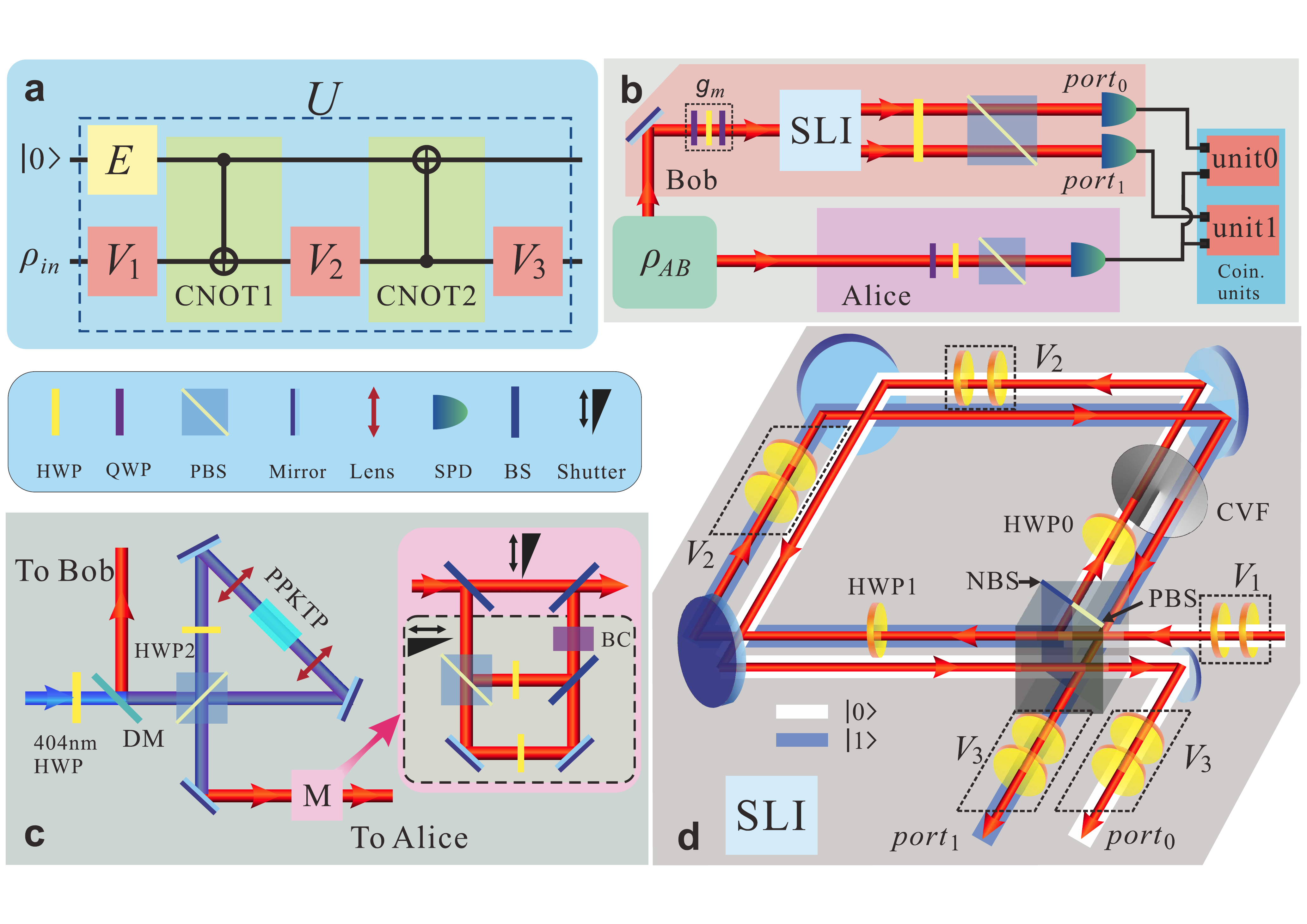}
\caption{Logic circuit and experimental setup. {\bf a} The logic circuit for implementing $U$. {\bf b} The integrated experimental setup. One photon is sent to Bob, and the other is sent to Alice. On Bob's side, each one of the gates $g_m$ before the Sagnac-like interferometer (SLI) is realized using a combination of a quarter-wave plate (QWP), a half-wave plate (HWP) and a QWP; the photons are measured along the $z$ direction using an HWP and a polarized beam splitter (PBS). On Alice's side, in the single-qubit protocol, the photons are detected directly to provide a coincidence signal. While, in the two-qubit protocol, Alice measures her photons along a direction $\vec{n}_i$ that is chosen based on the result $b|g_m$ received from Bob. The photons on both sides are detected by single-photon detectors (SPD). Finally, Alice's measurement result is sent to coincidence units, unit0 and unit1, to coincide with the corresponding results from $port_0$ and $port_1$, respectively. {\bf c} The unit used to prepare the investigated entangled states. The polarization Sagnac interferometer is used to prepare the maximally entangled state $|\Phi\rangle$ to be fed into the dual-wavelength PBS and HWP, i.e., HWP2. An additional unit M, in which the dashed gray part inserted with a long enough birefringent crystal (BC) assists in preparing the maximally mixed component $\mathbb{I}/4$, is placed at the port to Alice to produce the mixed state $\rho_{AB}$. Two moveable shutters are used to adjust the parameter $\eta$. BS, beam splitter; DM, dichroic mirror. {\bf d} Experimental realization of $U$ with the SLI constructed from a homemade beam splitter, with half of it coated as a PBS and the other half coated as a non-polarized beam splitter (NBS).}\label{setup}
\end{figure}

{\it Experimental setup.} The unitary operation $U$ shown in Fig. \ref{theory} {\bf c} can be decomposed into several parts, including two control-not (CNOT) gates ($C_{\text{NOT1}}$ and $C_{\text{NOT2}}$) and the other unitary evolutions $E$, $V_1$, $V_2$ and $V_3$, and implemented in an optical Sagnac-like interferometer (SLI), as illustrated in Fig. \ref{setup} (see Methods). And $U$ could be expressed as
\begin{equation}\label{uope}
U=(\mathbb{I}_C\otimes V_3)\cdot C_{\text{NOT2}}\cdot (\mathbb{I}_C\otimes V_2)\cdot C_{\text{NOT1}}\cdot (E\otimes V_1),
\end{equation}
where $\mathbb{I}_C$ is the identical operation on the control qubit. To obtain the bound $P^s$ and verify the setup, the single-qubit protocol is performed with the input state denoted as $\rho(\theta)=\cos\theta\,|H\rangle+\sin\theta\,|V\rangle$ where $|H\rangle$ and $|V\rangle$ represent the horizonal and vertical polarizations of the photons, respectively. For the two-qubit protocol, Werner states are prepared via the spontaneous parametric down conversion process by pumping the nonlinear crystal of periodically poled $\text{KTiOPO}_4$ (PPKTP) which is placed in a polarization Sagnac interferometer \cite{kim2006}. Here, $|\Phi\rangle$ is prepared to be $(|HH\rangle+|VV\rangle)/\sqrt{2}$. The experimental Werner states $\rho_{AB}$ are prepared with an average fidelity of $98.3\pm0.2\%$. The detailed experimental preparation can be found in Methods.

\begin{figure}[!h]
\centering
\includegraphics[width=0.47\textwidth]{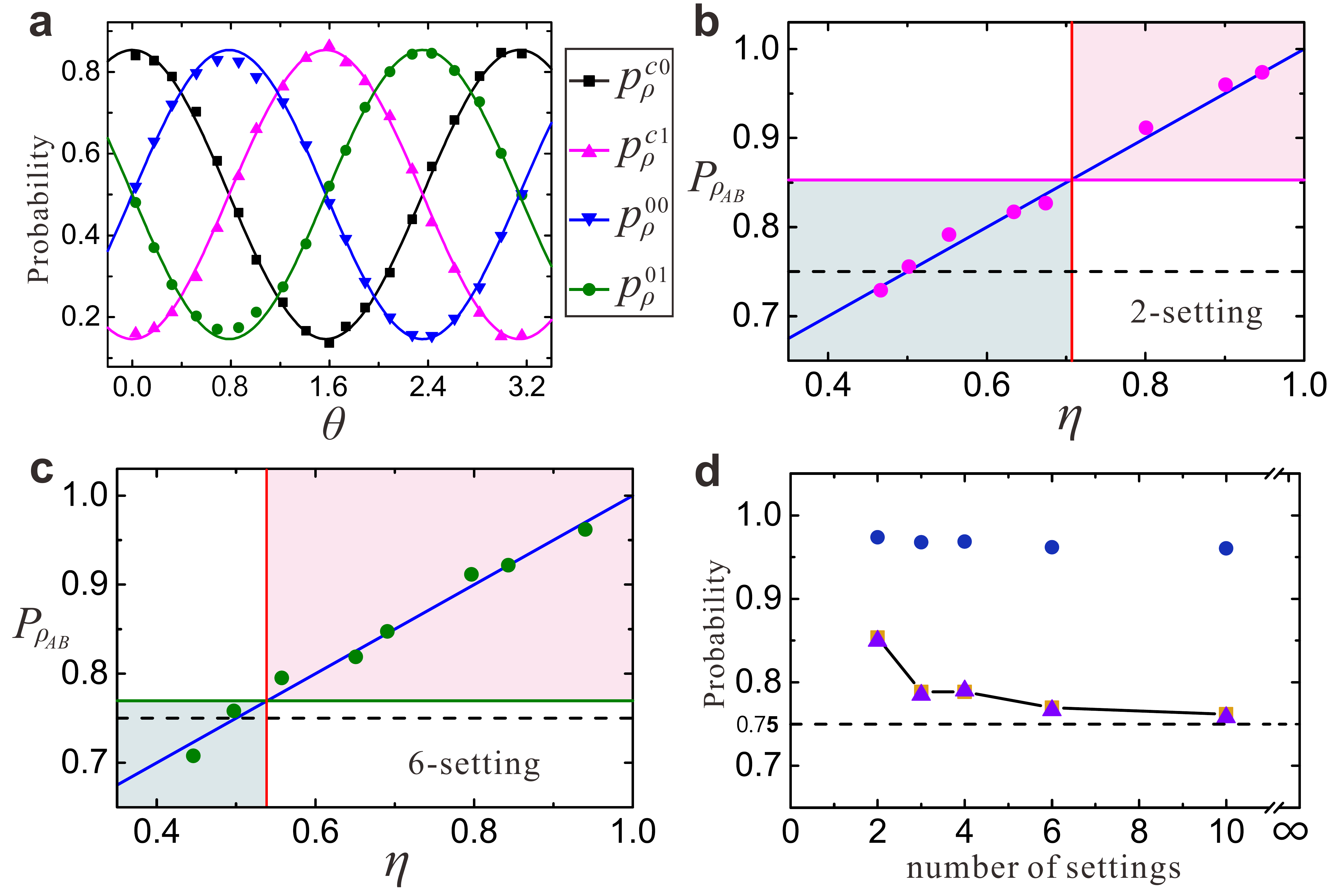}
\caption{Experimental results for the SD task. {\bf a} The probabilities of successful discrimination with single-qubit states in the case of two measurement settings. The curves and symbols represent the theoretical predictions and experimental results, respectively. {\bf b} and {\bf c} The experimental results in the two-qubit protocol with two and six measurement settings, respectively. The blue lines represent the theoretical predictions. The pink and green dots in {\bf b} and {\bf c} represent the corresponding experimental results, with the pink and green solid lines representing the single-qubit upper bounds for two and six measurement settings, respectively. The black dashed lines represent the single-qubit upper bound for the infinite number of measurement settings. {\bf d} The comparison of the single-qubit upper bounds $P^s$ with the results $P_{\rho_{AB}}$ obtained using the prepared maximally entangled state (blue dots) for different number of measurement settings. The success probabilities $P_{\rho_{AB}}$ are lower than the theoretical predictions, which can be primarily attributed to imperfect experimental manipulation. The brown squares represent the theoretical predictions of the upper bound for single-qubit states, whereas the experimental results are represented by the purple triangles. The bound $P^s$ established for the single-qubit approach decreases, and it is very close to the value with infinite measurement settings when the number of measurement setting is equal to ten. The experimental error bars which are very small and not shown are estimated as the standard deviation.}\label{result1}
\end{figure}

{\it Experimental results.} In the case of two measurement settings, the results $p_{\rho}^{c0},\,p_{\rho}^{c1},\,p_{\rho}^{00}$ and $p_{\rho}^{01}$ are presented in Fig. \ref{result1} {\bf a} which show that the input states $\rho(\theta)$ should be optimized to obtain the upper-bound value $P^s$. More results in the single-qubit protocol with multiple measurement settings are presented in Section III of Appendix. The Werner state $\rho_{AB}$ is identified to be steerable when the SD performance is enhanced with $P_{\rho_{AB}}>P^s$, see Fig. \ref{result1} {\bf b} and {\bf c}. By contrast, when $P_{\rho_{AB},\,n}\leq P_n^s$ ($n=2,\,3,\,4,\,6,\,10$), Alice fails to steer Bob's state via the corresponding SD task. As the number of measurement settings increases, the bound established for the single-qubit approach decreases, whereas the success probability achieved by employing steerable resources remains constant, as illustrated in Fig. \ref{result1} {\bf d}. All error bars in this work are estimated as the standard deviation from the statistical variation of the photon counts, which is assumed to follow a Poisson distribution. As the error bars on the experimental probabilities are very small, roughly $0.002$, they are not shown in the figures.

\begin{figure}[!h]
  \centering
  \includegraphics[width=0.47\textwidth]{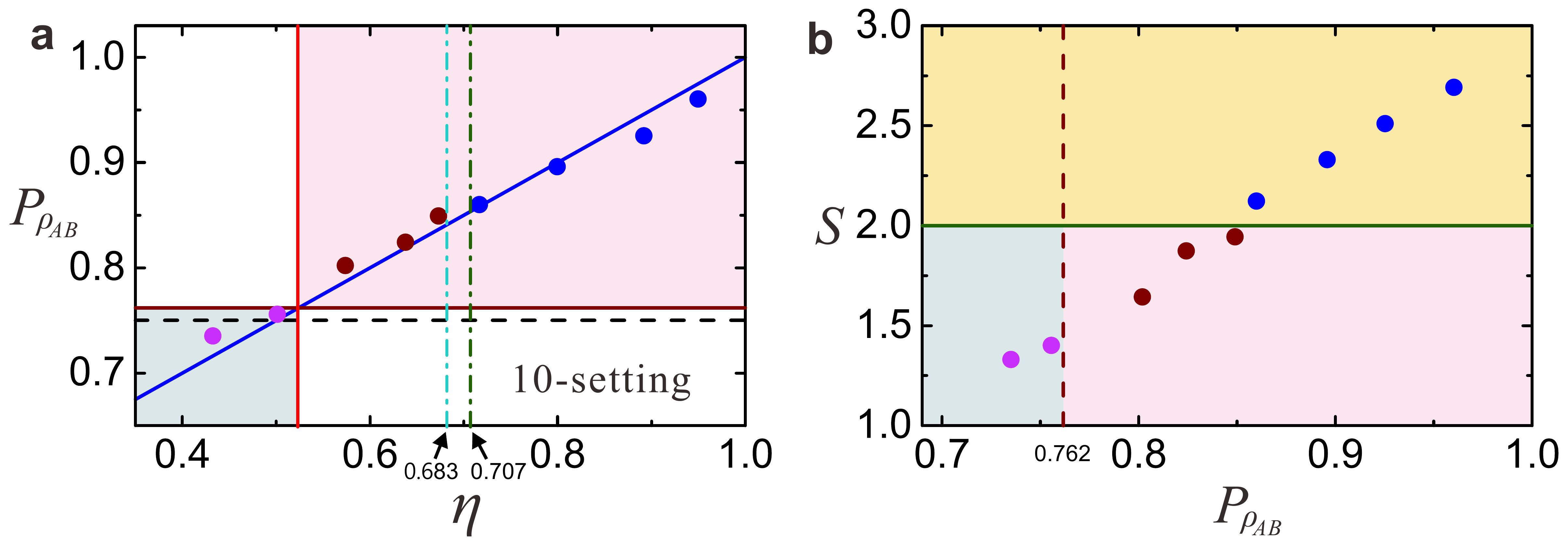}\\
  \caption{Experimental results of investigating different kinds of correlations via the SD task. {\bf a} The probabilities of successful discrimination for the case of ten measurement settings. The colored dots represent the experimental results, and the blue line represents the theoretical prediction. The dark red solid line and black dashed lines represent the upper bounds in the case of the single-qubit protocol for ten and infinite measurement settings, respectively. {\bf b} The experimental results for the Bell-CHSH parameter $S$ as a function of the success probability of SD, $P_{\rho_{AB}}$, in the case of ten measurement settings. The colored dots represent the experimental results. The green solid line represents the upper bound of local-hidden-variable model. The dark red dashed line represents the single-qubit upper bound. The experimental error bars which are very small and not shown are estimated as the standard deviation.
  }\label{result2}
\end{figure}

Moreover, by means of the SD task, the difference between EPR steering and entanglement can be characterized in an operational way. It is found that the success probability achieved using unsteerable Werner states $\rho_{AB}$ cannot surpass the single-qubit bound when $\eta\leq\eta_{10}^*\approx0.524$ in the case of ten measurement settings \cite{saunders2010}. However, $\rho_{AB}$ is still entangled when $\eta>1/3$. This implies that the success probability of SD cannot be enhanced by using unsteerable entangled states, which is experimentally verified by the two pink dots in Fig. \ref{result2} {\bf a}. The concurrences of these two pink dots are measured to be $0.154\pm0.008$ and $0.223\pm0.009$, which verify that the states are entangled \cite{wootters1998}. We further investigate EPR steering with Bell-local states. Theoretically, the Bell inequality will be violated when $\eta>1/\sqrt{2}$ \cite{werner1989}, and according to Ref. \cite{hirsch2016}, $\rho_{AB}$ is a Bell-local state when $\eta\leq0.683$. We measure the Bell-CHSH parameter $S$ \cite{chsh1969} which is shown as the function of $P_{\rho_{AB}}$ in Fig. \ref{result2} {\bf b}. The success probabilities of SD using three Bell-local states which are represented by the dark red dots in Fig. \ref{result2} are enhanced, and therefore, these states are steerable.

\section{Discussion}

Based on the proof of the necessary and sufficient characterization of EPR steering, we designed and experimentally implemented an SD task to demonstrate EPR steering using two-qubit Werner states. The methods for decomposing a quantum evolution into subchannels can be helpful for gaining a thorough understanding of complex open-system dynamics. The enhanced probabilities of successful discrimination achieved using EPR steering provides a concrete example of the application. Moreover, this practical task offers an intuitive means of operationally distinguishing the different concepts of quantum nonlocality.

Compared with the previous experiments using steering inequalities to investigate EPR steering, in which Bob measures along several directions when steered by Alice \cite{saunders2010,sun2016}, our work exhibits a particular feature that the measurement performed on Bob's qubit is restricted to a single direction which is $z$ in this work. This feature implies that the SD task offers a convenient approach for identifying EPR steering. Another character of the SD task is the measurement sequence of Alice and Bob. In the previous works \cite{wjd2007,sun2014}, considering that Alice steers Bob, Bob performs the measurements after receiving the measurement results from Alice. However, in the SD task, the sequence is reversed which means Alice begins to measure her qubit after Bob's measurements.

As EPR steering can be regarded as the one-side device-independent quantum information task \cite{cyril2012}, the steering-enhanced SD task, where Bob trusts his experimental device while Alice's side is device-independent, shows the potential application in one-side device-independent quantum key distribution. Furthermore, in our work, the SD task on Bob's side is implemented based on the one-way classical communication (from Bob to Alice). Considering the situation that Bell nonlocality relates to the two-side device-independent quantum information task \cite{jones2007,cyril2012}, one might extend the SD task demonstrating EPR steering to investigate the Bell nonlocality. For instance, a similar quantum information task referring to a bipartite SD problem (SD tasks on both sides) with two-way classical communications, which relates to the communication complexity problem \cite{zei2004,buhrman2016}, might be used to characterize Bell nonlocality in an operational way.

\section{Methods}

\subsection{The detailed expressions of $A_j$ and $K_{ij}$ in the two-setting case}

Following the theoretical method to determine $A_j$ which is introduced in Section I of Appendix, we can obtain the expressions of $A_0,\ A_1$ in the two-setting case as below
\begin{equation}\label{a0a1}
A_{0}=\left(\begin{array}{cc}
\frac{1}{4\sin{\frac{\pi}{8}}} & \frac{\sin{\frac{\pi}{8}}}{\sqrt{2}}\\
\frac{1}{4\sin{\frac{\pi}{8}}} & -\frac{\sin{\frac{\pi}{8}}}{\sqrt{2}}\\
\end{array}\right),\,\,\,\,
A_{1}=\left(\begin{array}{cc}
\frac{\sin{\frac{\pi}{8}}}{\sqrt{2}} & -\frac{1}{4\sin{\frac{\pi}{8}}}\\
\frac{\sin{\frac{\pi}{8}}}{\sqrt{2}} & \frac{1}{4\sin{\frac{\pi}{8}}}\\
\end{array}\right).
\end{equation}

Considering the Kraus operators $K_{ij}$ which satisfy $K_{i,j}=|i\rangle\langle i|\cdot A_j$ ($i,\,j=0$ or $1$), we can get
\begin{equation}\label{kraus}
\begin{split}
K_{00}=\left(\begin{array}{cc}
\frac{1}{4\sin{\frac{\pi}{8}}} & \frac{\sin{\frac{\pi}{8}}}{\sqrt{2}}\\
0 & 0\\
\end{array}\right),\,\,\,
K_{01}=\left(\begin{array}{cc}
\frac{\sin{\frac{\pi}{8}}}{\sqrt{2}} & -\frac{1}{4\sin{\frac{\pi}{8}}}\\
0 & 0\\
\end{array}\right),\\
K_{10}=\left(\begin{array}{cc}
0 & 0\\
\frac{1}{4\sin{\frac{\pi}{8}}} & -\frac{\sin{\frac{\pi}{8}}}{\sqrt{2}}\\
\end{array}\right),\,\,\,\,
K_{11}=\left(\begin{array}{cc}
0 & 0\\
\frac{\sin{\frac{\pi}{8}}}{\sqrt{2}} & \frac{1}{4\sin{\frac{\pi}{8}}}\\
\end{array}\right).
\end{split}
\end{equation}

For multi-setting cases, the corresponding expressions of $A_j$ and $K_{ij}$ can be obtained using the similar method which is shown in Section I of Appendix.

\subsection{Experimental implementation of the unitary operation $U$}

We construct an inherently stable optical interferometer, namely, a Sagnac-like interferometer (SLI), to realize this operation $U$ (see Fig. \ref{setup} {\bf b}). The path and polarization degree of freedom of the photons are used as the auxiliary qubit, which is initially in the state $|0\rangle$, and the probe qubit, respectively. A homemade beam splitter, of which one half is coated as a PBS and the other half is coated as a non-polarized beam splitter (NBS), acts as the input-output coupling element of the interferometer. Each single-qubit gate evolution of the probe qubit (the polarization of photons), i.e., $V_1,\,V_2$, and $V_3$, is realized through a combination of two HWPs. The operation $E$ on the auxiliary qubit is realized by adjusting the ratio of the numbers of photons on the $|0\rangle$ and $|1\rangle$ paths, which is achieved by means of a continuously variable neutral density filter (CVF) crossing both paths. For the first CNOT gate, the path qubit is the control qubit, while the polarization is used as the target qubit. Thus, the polarization of photons on the $|0\rangle$ path remains the same, whereas the polarization on the $|1\rangle$ path reverses, meaning that the polarization $|H\rangle$ is flipped to $|V\rangle$ and $|V\rangle$ is flipped to $|H\rangle$. This process is realized by placing one HWP on each of the two paths; HWP0, located on the $|0\rangle$ path, is set at $0^\circ$ for phase compensation, while HWP1 is set at $45^\circ$ to reverse the polarizations of $|H\rangle$ and $|V\rangle$. The second CNOT gate is the inverse of the first CNOT gate; the polarization is treated as the control qubit affecting the target qubit, which is the qubit related to the path information. This gate is implemented in the PBS part of the homemade beam splitter. In detail, the $|H\rangle$ polarization remains unchanged (retaining the same path information), while the $|V\rangle$ polarization flips to the other path. The imperfect optical elements, especially the homemade beam splitter, would reduce the visibility of the interferometer and introduce system errors.

To realize $\{g_m\}$ in the multi-setting cases, several wave plates including HWPs and quarter-wave plates (QWPs) are employed. This part is explained in detail in Section II of Appendix.

\subsection{Preparation of the experimental states}

To obtain the single-qubit upper bound and verify the setup, we perform the SD task using the following single-qubit state $\rho(\theta)$,
\begin{equation}\label{sta1}
  \rho(\theta)=\cos\theta|H\rangle+\sin\theta|V\rangle.
\end{equation}
In this case, the photons on Bob's side are prepared as the state expressed in Eq. \ref{sta1}, and the photons on Alice's side are detected directly to provide coincidence signals. $\rho(\theta)$ are simply prepared with a half-wave plate (HWP) set at the angel $\theta/2$ following a polarized beam splitter (PBS). The upper bound is then $P^s=\max_{\rho(\theta)}\{P^s_{\rho(\theta)}\}$.

The investigated $\rho_{AB}$ states are manufactured by combining the maximally entangled state $|\Phi\rangle$ and the maximally mixed state $\mathbb{I}/4$. $|\Phi\rangle$ is prepared via the spontaneous parametric down conversion process where a $\chi^{(2)}$ nonlinear crystal of periodically poled $\text{KTiOPO}_4$ (PPKTP) is pumped by an ultraviolet laser with a peak wavelength of $404.1\, nm$ and a spectrum width of $0.05\,nm$. The crystal is placed in a polarization Sagnac interferometer \cite{kim2006}, as illustrated in Fig. \ref{setup} {\bf c}. The dichroic mirror (DM) is designed to exhibit high transmission at $404\, nm$ and high reflection at $808\, nm$. A dual-wavelength polarization beam splitter (PBS) is employed as the input-output coupling element of the Sagnac interferometer, and a dual-wavelength HWP set at $45^\circ$ is used to change the vertically polarized component of the ultraviolate photon to the horizonal polarization to pump the PPKTP crystal. The crystal is placed in a thermoelectric oven with the temperature set at $28.5\pm0.1^\circ$C. The maximally entangled state $|\Phi\rangle$ is prepared with a brightness of $\sim18000$ $\text{pairs}\cdot \text{s}^{-1}\cdot \text{mW}^{-1}$, which is filtered using $3\, nm$ bandwidth filters, and the state fidelity is $95.5\pm0.4\%$. As shown in Fig. \ref{setup} {\bf c}, a part of the input of unit M still remains as the maximally entangled state $|\Phi\rangle\langle\Phi|$, and the other part is used to prepare the maximally mixed state $\mathbb{I}/4$ with the dashed gray part in unit M. Two HWPs are set at $22.5^\circ$ and a birefringent calcite crystal (BC) of $10\,mm$ in length is employed to induce decoherence between the horizonal and vertical polarizations of the photons. The shutters are used to adjust the ratio between $|\Phi\rangle\langle\Phi|$ and $\mathbb{I}/4$ to control the parameter $\eta$.

\section{Appendix}
\section{The method to determine $A_j$, $g_m$ and the guessing strategies}

This section will contain three topics.
\begin{itemize}
\item Firstly, we will provide a heuristic approach to find the appropriate $A_j$ and $g_m$ for the SD task with Werner states. It is not a strict proof but just an intuitive guiding which could also help us to find the corresponding SD task for other states.
\item Then we will give the proofs. Once we find $A_j$ and $g_m$, the proof of their suitableness is relatively simple. We could directly calculate the upper-bound successful probability of the SD task $P^s_{n}$ under $n$ measurement settings given by the single qubit and the probability $P_{\rho_{AB},\,n}$ given by the Werner state. If $P_{\rho_{AB},\,n}> P_n^s$, the steerability of $\rho_{AB}$ is certified.
\item Lastly, we will discuss the ability of the constructed SD task to test the EPR steering. To improve the testing ability, we need to decrease the successful probability of the single-qubit protocol. Since the linear steering inequalities \cite{wjd2007} is a pretty good criteria for Werner states, we need $P^{s}_{n}$ match the bound of $C_n^{LHS}$, and here, $P_n^s=1/2+\eta_n^*/2=1/2+C_n^{LHS}/2$. In this work, we consider the regularly spaced measurement settings given by the Platonic solids. In some cases ($n=4$ and $6$), it has been proven the regula measurements are suboptimal choices \cite{evans2014}. We endured this defect with two excuses, (i) the testing ability improved by the optimal settings are limited; (ii) the corresponding SD task given by the irregular multi-setting measurement is difficult to experimentally realize. In our work, since the multi-setting is realized by adding a group of $\{g_m\}$ before the operation $K_{ij}$ instead of increasing the dimension of Hilbert space of the auxiliary system, the experiment difficulties of the multi-setting cases are the same with the two-setting case. While, this facility way is blocked for the irregularly spaced measurement. The detail will be discussed later.
\end{itemize}

In this work, for every case of $n$ ($n=2,\,3,\,4,\,6,\,10$) measurement settings, during the designs of subchannels ($A_j$ and $g_m$) and strategies, we follow these two rules (i) the success probability of maximally entangled state is 1; (ii) The success probability of maximally mixed state is 1/2. Thus for each Werner state $\rho_{AB}$ shown in Eq. (2) in the main text, the success probability equals to $P_{\rho_{AB}}=1\cdot\eta + 1/2 \cdot (1-\eta)=1/2+\eta/2$. Once we follow these two rules, the success probability of SD task given by the Werner state is fixed. Thus to improve the ability of testing steering we need decrease the minimum probability $P^s_{n}$ given by the single-qubit states.

\subsection{The method for the two-setting case}

Let us discuss the two-setting case first. In this case two orthogonal directions are optimal choice for testing steering for the Werner state. The measurement settings $\vec{n}_i$ on Alice's side could be written as $\vec{n}_b$, which means $\vec{n}_i$ are chosen based on the result $b$. Without lose of generality, let us chosen as $\vec{n}_0=(1/\sqrt{2},0,1/\sqrt{2})$ and $\vec{n}_1=(-1/\sqrt{2},0,1/\sqrt{2})$. The success probability of SD by employing $\rho_{AB}$ assisted with result of $b$ \cite{piani2015} is
\begin{equation}\label{si1}
P_{\rho_{AB}}=\sum_{j,b}\text{Tr}[\Pi_b^B\cdot(A_j\cdot P_{j|\vec{n}_b}^B \rho_{j|\vec{n}_b}^B\cdot A_j^\dag)],
\end{equation} where the corresponding strategy is that Alice measures along $\vec{n}_b$ and gets the result $a$, and then guess $j=a$. Here $\Pi_b^B$ is the POVM on Bob's side, and because the medium subchannel $A_j$ is followed by an entanglement-breaking channel (EBC) (See Fig. 1 (b) in the main text), the optimal measurement setting by Bob is along $\vec{n}=z$, i.e., $\Pi_b^B=\frac{\mathbb{I}+(-1)^b\sigma_z}{2}$; $\rho_{j|\vec{n}_b}^B$ is the normalized conditional state on Bob's state when Alice gets result $j$ measuring along $\vec{n}_b$, $P_{j|\vec{n}_b}^B$ is the corresponding probability. Eq. \ref{si1} can be rewriten as:
\begin{equation}\label{si2}
P_{\rho_{AB}}=\sum_{j,b}\text{Tr}[(A_j^\dag\cdot\Pi_b^B\cdot A_j)\cdot P_{j|\vec{n}_b}^B \rho_{j|\vec{n}_b}^B].
\end{equation}

\begin{figure}[!h]
\centering
\includegraphics[width=0.4\textwidth]{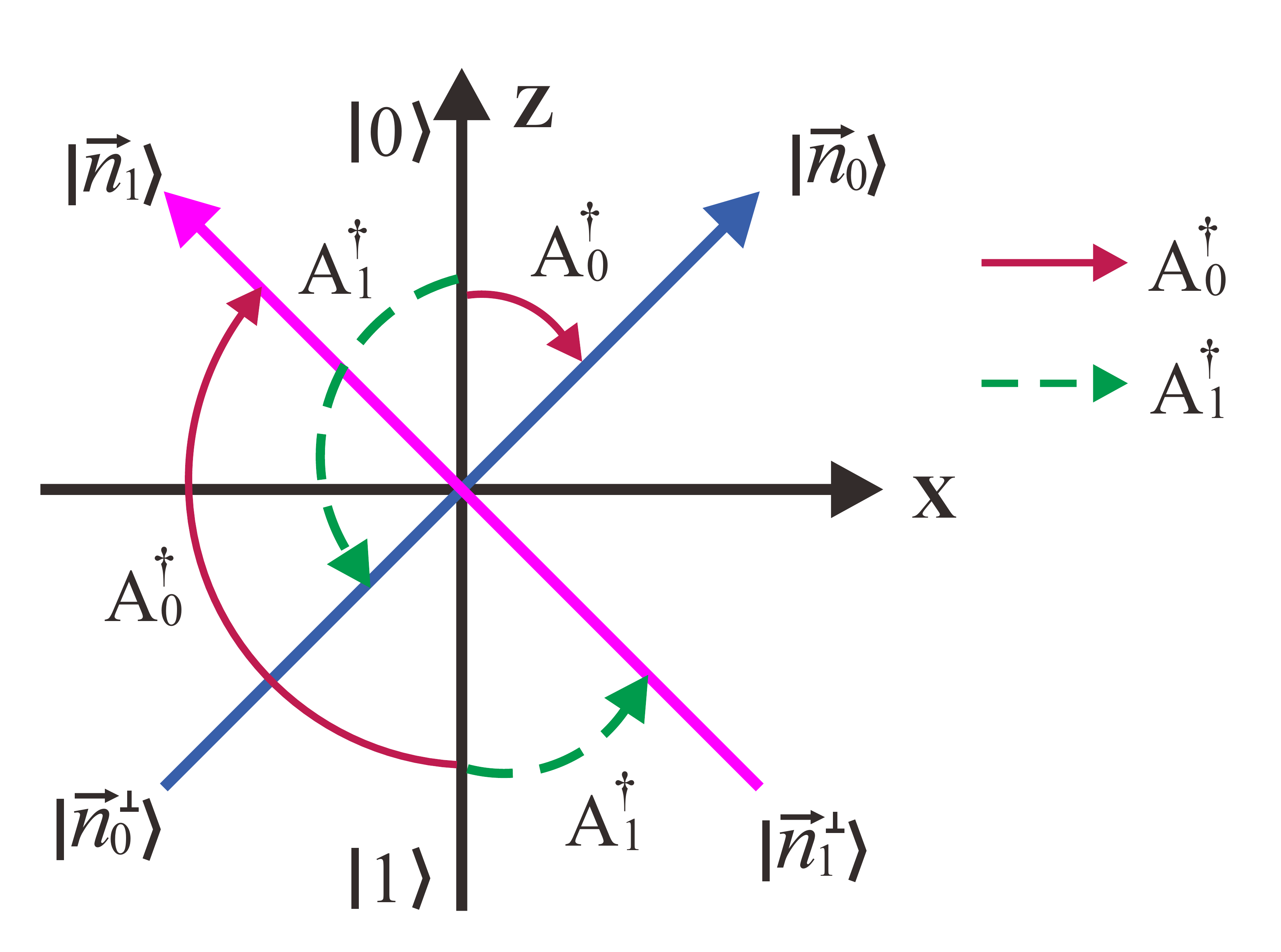}
\caption{Illustrating the functions of $A_0^\dag$ and $A_1^\dag$.
}\label{figs1}
\end{figure}

Let us consider the case $\rho_{AB}$ is the maximally entangled state in which $\rho_{j|\vec{n}_b}^B$ is a pure state. This form can also be regarded as the situation where the quantum state $\Pi_b^B$ denoted by $|b\rangle\langle b|$ ($b=0$ or $1$) is followed by a subchannel $A_j^\dag$ and then measured along $\vec{n}_b$. As shown in Fig. \ref{figs1}, to maximize the discrimination probability with the condition $|0\rangle$ or $|1\rangle$, $A_0^\dag$ and $A_1^\dag$ are designed to satisfy that $A_0^\dag$ turns $|0\rangle$ to $|\vec{n}_0\rangle$ and $A_1^\dag$ turns $|0\rangle$ to $|\vec{n}_0^\bot\rangle$, where $|\vec{n}_0\rangle$ and $|\vec{n}_0^\bot\rangle$ are the two eigenstates of the direction $\vec{n}_0$. Thus in the case of $|0\rangle$, when Alice measures along $\vec{n}_0$, $|\vec{n}_0\rangle$ and $|\vec{n}_0^\bot\rangle$ are perfectly distinguishable. Similarly in the case of $|1\rangle$, $A_0^\dag$ turns $|1\rangle$ to $|\vec{n}_1\rangle$ and $A_1^\dag$ turns $|1\rangle$ to $|\vec{n}_1^\bot\rangle$. In this case, for the maximally entangled state $|\Psi\rangle$, the discrimination probability is unit. We have
\begin{equation}\label{lhs}
\left\{
\begin{aligned}
& A_0^\dag \cdot |0\rangle\langle0| \cdot A_0 \varpropto |\vec{n}_0\rangle\langle\vec{n}_0| \\
& A_1^\dag \cdot |0\rangle\langle0| \cdot A_1 \varpropto |\vec{n}_0^\bot\rangle\langle\vec{n}_0^\bot| \\
& A_0^\dag \cdot |1\rangle\langle1| \cdot A_0 \varpropto |\vec{n}_1\rangle\langle\vec{n}_1| \\
& A_1^\dag \cdot |1\rangle\langle1| \cdot A_1 \varpropto |\vec{n}_1^\bot\rangle\langle\vec{n}_1^\bot| \\
\end{aligned}
\right.
\end{equation}

Considering that $\{A_j^\dag\}_j$ is a group of subchannels, we get:
\begin{equation}\label{subc2}
A_{0}=\left(\begin{array}{cc}
\frac{1}{4\sin{\frac{\pi}{8}}} & \frac{\sin{\frac{\pi}{8}}}{\sqrt{2}}\\
\frac{1}{4\sin{\frac{\pi}{8}}} & -\frac{\sin{\frac{\pi}{8}}}{\sqrt{2}}\\
\end{array}\right),\,\,\,\,
A_{1}=\left(\begin{array}{cc}
\frac{\sin{\frac{\pi}{8}}}{\sqrt{2}} & -\frac{1}{4\sin{\frac{\pi}{8}}}\\
\frac{\sin{\frac{\pi}{8}}}{\sqrt{2}} & \frac{1}{4\sin{\frac{\pi}{8}}}\\
\end{array}\right).
\end{equation}

Note that the information of $i$ is given by the output of $b$, and to discriminate $K_{ij}$ is equivalent to guess $j$. For Werner states, the optimal guessing strategy is guessing $j=a$ with $a$ obtained when Alice measures along $\vec{n}_b$. By directly calculating, we get $P_{\rho_{AB}}=1/2+\eta/2$. Now let us calculate the maximum probability of the single-qubit protocol. There are four types of strategies for the single-qubit resource: (i) always guessing $j=0$ and denoting the success probability as $p_{\rho}^{c0}$ with the input state $\rho$, in this case the optimal input state is $\rho=|0\rangle$; (ii) always guessing $j=1$ and denoting the success probability as $p_{\rho}^{c1}$ with $\rho$, the corresponding optimal input state is $\rho=|1\rangle$; (iii) guessing $j=b$ and denoting the success probability as $p_{\rho}^{00}$ with $\rho$, and the optimal input state is $\rho=|+\rangle=(|0\rangle+|1\rangle)/\sqrt{2}$; and (iv) guessing $j=b\oplus1$ and denoting the success probability as $p_{\rho}^{01}$ with $\rho$, the optimal input state is $\rho=|-\rangle=(|0\rangle-|1\rangle)/\sqrt{2}$. For a given input state $\rho$, the success probability is $P^s_{\rho}=\max{\{p_{\rho}^{c0},\,p_{\rho}^{c1},\,p_{\rho}^{00},\,p_{\rho}^{01}\}}$, and the maximum probability is $P^s=\max_{\rho}{\{P^s_{\rho}\}}$. Through direct calculation, we get $P^s_{2}=(1+1/\sqrt{2})/2=1/2+\eta_2^*/2$ where $\eta_2^*$ is the visibility bound of the Werner states in the case of two measurement settings. When $\eta>\eta_2^*$, $\rho_{AB}$ is steerable.

\subsection{The method for the multi-setting cases}

In the multi-setting cases, the measurement settings on Alice's side are chosen to be the directions along the antipodal pairs of vertices of the Platonic solids \cite{saunders2010}. We need to determine both $A_j$ and $g_m$. Taking four measurement settings for example, we need two quantum gates $g_m$, i.e., $g_1$ and $g_2$, and Alice's measurement settings are along the antipodal pairs of vertexes of a cube (See Fig. \ref{34setting} (d)). In this case, we could choose $\vec{n}_0=(\sqrt{\frac{2}{3}},0\sqrt{\frac{1}{3}})$, $\vec{n}_1=(-\sqrt{\frac{2}{3}},0\sqrt{\frac{1}{3}})$ for the quantum gate $g_1=\mathbb{I}$ and $g_1$ is always identical operation in other multi-setting cases. For the quantum gate $g_2$ where the evolution is rotating along $z$ with $\pi/2$, i.e., $g_2=\bigl(\begin{smallmatrix} 1 & 0 \\ 0 & -i \end{smallmatrix}\bigr)$, $\vec{n}_0$ and $\vec{n}_1$ will be rotated to the other two directions in the cube. Due to the fact that which one of the quantum gates $g_m$ is performed is unknown before the device implementing the evolution $K_{ij}$ (See Fig. 1 (d) in the main text), the maximum success probability in the single-qubit protocol $P^s_n$ is reduced. On the other hand, in the two-qubit protocol, both the measurement result $b$ and the information $g_m$ on Bob's side will be sent to Alice, and this does not reduce the success probability of the two-qubit state. For example, the maximally entangled state $|\Psi\rangle$, in the multi-setting case, the success probability of SD is still $100\%$ and the success probability of Werner state is $P_{\rho_{AB}}=1/2+\eta/2$.

Now let us calculate the maximum success probability of single-qubit state $P^s_4$ in the case of four measurement settings. Once sending the single qubit into the subchannel and getting the output, we can obtain the information about $b$ and $m$ (we don't know which $m$ is selected before his input). The best strategy of guessing $j$ based on $b$ and $m$ is shown in Table \ref{stra4}.

\begin{table*}[htbp]
\renewcommand\arraystretch{1}
  \centering
  \begin{tabular}{|p{4cm}<{\centering}|p{4cm}<{\centering}|p{4cm}<{\centering}|}\hline
  Guessing Strategy  & Optimal Input State & Success Probability \\\hline
  $(0,\,0)$ & $(0,\,0,\,1)$ & $(1+1/\sqrt{3})/2$ \\\hline
  $(1,\,1)$ & $(0,\,0,\,-1)$ & $(1+1/\sqrt{3})/2$ \\\hline
  $(b,\,b)$ & $(1/\sqrt{2},\,1/\sqrt{2},\,0)$ & $(1+1/\sqrt{3})/2$ \\\hline
  $(b\oplus 1,\,b\oplus 1)$ & $(-1/\sqrt{2},\,-1/\sqrt{2},\,0)$ & $(1+1/\sqrt{3})/2$ \\\hline
  $(b,\,b\oplus 1)$ & $(1/\sqrt{2},\,-1/\sqrt{2},\,0)$ & $(1+1/\sqrt{3})/2$ \\\hline
  $(b\oplus 1,\,b)$ & $(-1/\sqrt{2},\,1/\sqrt{2},\,0)$ & $(1+1/\sqrt{3})/2$ \\\hline
\end{tabular}
  \caption{The guessing strategy using the single-qubit state in the four-setting case. $b\in \{0,1\}$, $\oplus$ means addition modulo $2$ and the input states are expressed by a unit direction $\vec{n}_{\rho}$ in the Bloch sphere. And $\rho=(\mathbb{I}+\vec{\sigma}\cdot \vec{n}_{\rho})/2$ where $\vec{\sigma}=(\sigma_x,\sigma_y,\sigma_z)$ is the Pauli vector. The column of guessing strategy has two terms which correspond to different $m=1,\,2$. For example, in the last row, ($b\oplus 1$,\,b) means guessing $j=b\oplus 1$ if $m=1$, and guessing $j=b$ if $m=2$. The column of success probability shows the maximum probability of guessing $j$ by using the optimal input state. For each $m$, the corresponding success probability is $(1+1/\sqrt{3})/2$.}\label{stra4}
\end{table*}

In the case of testing steering with linear steering inequalities, if Bob's measurement settings are vertexes of the cube for 4-setting, we can certify the steerability of Werner states with a visibility higher than $C^{LHS}_4=1/\sqrt{3}$. The testing ability is the same with our SD task where $P^s_4=(1+1/\sqrt{3})/2$. As Ref. \cite{evans2014} shows this testing ability could be improved with irregular spaced measurement settings consisting of three directions regularly spaced in the $X-Y$ plane and the forth along direction $Z$. These settings could test steering with a visibility higher than $C^{opt}_4=2/\sqrt{13} \approx 0.555$ which is better than here $C^{LHS}_4 \approx 0.577$. The improvement presented in the succuss probability is $(C^{LHS}_4-C^{opt}_4)/2 \approx 0.011$. While, normally speaking, a multi-setting steering test is corresponding to a multi-subchannel discrimination problem. According to the Stinespring dilation theorem \cite{stinespring1955}, such a problem needs a high dimension auxiliary system which is difficult to realize in experiment. In our work, to remove this obstruction we take the advantage of regular spaced measurement directions given by the Platonic solid. To increase the setting we just need to add a group of unity gates $\{g_m\}$ before the 2-subchannel discrimination problem. This is why we take the suboptimal choice in our experiment (for $n=4$ and $6$).

The results of other multiple settings are similar. By direct calculation we can get $P^s_n$. Compared with the bound $C^{LHS}_n$ given by the linear steering inequalities which measured with regular spaced directions, we find $P^s_{n}=(1+C^{LHS}_n)/2$, ($n=2,\,3,\,4,\,6,\,10$), and the details are shown below.

\subsection{The detailed expressions}

The $\{A_{j}\}$ or correspondingly the overall gate operation $U$ (see Fig.2 (a) in the main text) can be written as
\begin{equation}\label{cnot2}
U=(\mathbb{I}_C\otimes V_3)\cdot C_{\text{NOT2}}\cdot (\mathbb{I}_C\otimes V_2)\cdot C_{\text{NOT1}}\cdot (E\otimes V_1),
\end{equation}
where $\mathbb{I}_C$ is the identical operation on the control qubit. For different measurement settings $n$, the only differences are $\{g_m\}$ and $E$, where
\begin{equation}\label{cnot1}
\begin{split}
C_{\text{NOT1}}=\left(
\begin{array}{cccc}
1 & 0 & 0 & 0\\
0 & 1 & 0 & 0\\
0 & 0 & 0 & 1\\
0 & 0 & 1 & 0\\
\end{array}
\right),
C_{\text{NOT2}}=\left(
\begin{array}{cccc}
1 & 0 & 0 & 0\\
0 & 0 & 0 & 1\\
0 & 0 & 1 & 0\\
0 & 1 & 0 & 0\\
\end{array}
\right),
\end{split}
\end{equation}

and

\begin{equation}\label{logiv}
\begin{split}
V_1=\left(
\begin{array}{cc}
\frac{1}{\sqrt{2}} & -\frac{1}{\sqrt{2}}\\
\frac{1}{\sqrt{2}} & \frac{1}{\sqrt{2}}\\
\end{array}
\right),\,\,\,
V_2=\left(
\begin{array}{cc}
\frac{1}{\sqrt{2}} & \frac{1}{\sqrt{2}}\\
-\frac{1}{\sqrt{2}} & \frac{1}{\sqrt{2}}\\
\end{array}
\right),\\
V_3=\left(
\begin{array}{cc}
\frac{1}{\sqrt{2}} & -\frac{1}{\sqrt{2}}\\
\frac{1}{\sqrt{2}} & \frac{1}{\sqrt{2}}\\
\end{array}
\right).
\end{split}
\end{equation}

The details for different measurement settings $n$ are organized as followed. And according to $U=\bigl(\begin{smallmatrix} A_0 & -A_1 \\ A_1 & A_0 \end{smallmatrix}\bigr)$, and $U=(\mathbb{I}_C\otimes V_3)\cdot C_{\text{NOT2}}\cdot (\mathbb{I}_C\otimes V_2)\cdot C_{\text{NOT1}}\cdot (E\otimes V_1)$, we can obtain the corresponding expressions of $A_j$ and $K_{ij}$ in the multi-setting cases after getting the corresponding $E$.

$\bullet$ For $n=2$,
\begin{equation}\label{logic}
\begin{split}
E=\left(
\begin{array}{cc}
\cos \pi/8 & -\sin \pi/8\\
\sin \pi/8 & \cos \pi/8\\
\end{array}
\right).
\end{split}
\end{equation}
The single-qubit strategies are introduced in Table \ref{stra2}.

\begin{table*}[htbp]
\renewcommand\arraystretch{1}
  \centering
  \begin{tabular}{|p{4cm}<{\centering}|p{4cm}<{\centering}|p{4cm}<{\centering}|}\hline
  Guessing Strategy  & Optimal Input State & Success Probability \\\hline
  $j=0$ & $(0,\,0,\,1)$ & $(1+1/\sqrt{2})/2$ \\\hline
  $j=1$ & $(0,\,0,\,-1)$ & $(1+1/\sqrt{2})/2$ \\\hline
  $j=b$ & $(1,\,0,\,0)$ & $(1+1/\sqrt{2})/2$ \\\hline
  $j=b\oplus 1$ & $(-1,\,0,\,0)$ & $(1+1/\sqrt{2})/2$ \\\hline
\end{tabular}
  \caption{The guessing strategy using the single-qubit state in the two-setting case. The optimal input state is expressed by the same method in Table \ref{stra4}.}\label{stra2}
\end{table*}

The details of Alice's measurement settings when using the Werner state in the two-setting case are introduced in Table \ref{tstra2}. For $n=2,\,3,\,4,\,6,\,10$, Alice's measurement directions are based on $b|g_m$, the strategy is guessing $j=a$ and the success probability is $\frac{1}{2}+\frac{1}{2}\eta$, regardless $n$.

\begin{table*}[htbp]
\renewcommand\arraystretch{1}
  \centering
  \begin{tabular}{|p{4cm}<{\centering}|p{4cm}<{\centering}|p{4cm}<{\centering}|}\hline
  $b$  & Alice's Measurements  \\\hline
  $b=0$ & $(1/\sqrt{2},\,0,\,1/\sqrt{2})$  \\\hline
  $b=1$ & $(-1/\sqrt{2},\,0,\,1/\sqrt{2})$  \\\hline
\end{tabular}
  \caption{For Werner states, Alice's measurement directions are chosen based on $b$ in the two-setting case.}\label{tstra2}
\end{table*}

$\bullet$ For $n=3$ (see Fig. \ref{34setting} (a)),
\begin{equation}\label{logic3}
\begin{split}
&E=\left(
\begin{array}{cc}
\cos \pi/8 & -\sin \pi/8\\
\sin \pi/8 & \cos \pi/8\\
\end{array}
\right),
g_1=\left(
\begin{array}{cc}
1 & 0\\
0 & 1\\
\end{array}
\right),\\
&g_2=\left(
\begin{array}{cc}
\frac{i+\sqrt{2}}{2} & i/2\\
i/2 & \frac{-i+\sqrt{2}}{2}\\
\end{array}
\right),
g_3=\left(
\begin{array}{cc}
\frac{i-\sqrt{2}}{2} & -i/2\\
-i/2 & -\frac{i+\sqrt{2}}{2}\\
\end{array}
\right).
\end{split}
\end{equation}
The single-qubit strategies are shown in Table \ref{stra3}.

\begin{table*}[htbp]
\renewcommand\arraystretch{1}
  \centering
  \begin{tabular}{|p{4cm}<{\centering}|p{4cm}<{\centering}|p{4cm}<{\centering}|}\hline
  Guessing Strategy  & Optimal Input State & Success Probability \\\hline
  $(0,\,0,\,0)$ & $(0,\,-1/\sqrt{3},\,\sqrt{2}/\sqrt{3})$ & $(1+1/\sqrt{3})/2$ \\\hline
  $(1,\,1,\,1)$ & $(0,\,1/\sqrt{3},\,-\sqrt{2}/\sqrt{3})$ & $(1+1/\sqrt{3})/2$ \\\hline
  $(0,\,b,\,b \oplus 1)$ & $(0,\,1/\sqrt{3},\,\sqrt{2}/\sqrt{3})$ & $(1+1/\sqrt{3})/2$ \\\hline
  $(1,\,b\oplus 1,\,b )$ & $(0,\,-1/\sqrt{3},\,-\sqrt{2}/\sqrt{3})$ & $(1+1/\sqrt{3})/2$ \\\hline
  $(b,\,b,\,1 )$ & $(\sqrt{2}/\sqrt{3},\,1/\sqrt{3},\,0)$ & $(1+1/\sqrt{3})/2$ \\\hline
  $(b\oplus 1,\,b\oplus 1,\,0 )$ & $(-\sqrt{2}/\sqrt{3},\,-1/\sqrt{3},\,0)$ & $(1+1/\sqrt{3})/2$ \\\hline
  $(b,\,0,\,b )$ & $(\sqrt{2}/\sqrt{3},\,-1/\sqrt{3},\,0)$ & $(1+1/\sqrt{3})/2$ \\\hline
  $(b\oplus 1,\,1,\,b \oplus 1)$ & $(-\sqrt{2}/\sqrt{3},\,1/\sqrt{3},\,0)$ & $(1+1/\sqrt{3})/2$ \\\hline
\end{tabular}
  \caption{The guessing strategy using the single-qubit state in the three-setting case. The expressions in this table are the same with Table \ref{stra4}.}\label{stra3}
\end{table*}

The details of Alice's measurement settings when using the Werner state in the three-setting case are introduced in Table \ref{tstra3}.

\begin{table*}[htbp]
\renewcommand\arraystretch{1}
  \centering
  \begin{tabular}{|p{4cm}<{\centering}|p{4cm}<{\centering}|p{4cm}<{\centering}|}\hline
  $b|g_m$  & Alice's Measurements  \\\hline
  $0|g_1$ & $(1/\sqrt{2},\,0,\,1/\sqrt{2})$  \\\hline
  $0|g_2$ & $(1/\sqrt{2},\,0,\,1/\sqrt{2})$  \\\hline
  $0|g_3$ & $(0,\,1,\,0)$  \\\hline
  $1|g_1$ & $(-1/\sqrt{2},\,0,\,1/\sqrt{2})$  \\\hline
  $1|g_2$ & $(0,\,1,\,0)$  \\\hline
  $1|g_3$ & $(-1/\sqrt{2},\,0,\,1/\sqrt{2})$  \\\hline
\end{tabular}
  \caption{For Werner states, Alice's measurement directions are chosen based on $b|g_m$ in the three-setting case.}\label{tstra3}
\end{table*}

$\bullet$ For $n=4$ (see Fig. \ref{34setting} (d)),
\begin{equation}\label{logic}
\begin{split}
&E=\left(
\begin{array}{cc}
\sqrt{\frac{3+\sqrt{3}}{6}} & -\sqrt{\frac{3-\sqrt{3}}{6}}\\
\sqrt{\frac{3-\sqrt{3}}{6}} & \sqrt{\frac{3+\sqrt{3}}{6}}\\
\end{array}
\right),\\
&g_1=\left(
\begin{array}{cc}
1 & 0\\
0 & 1\\
\end{array}
\right),
g_2=\left(
\begin{array}{cc}
1 & 0\\
0 & -i\\
\end{array}
\right).
\end{split}
\end{equation}
The single-qubit strategies are shown in Table \ref{tstra4}.

The details of Alice's measurement settings when using the Werner state in the four-setting case are introduced in Table \ref{tstra4}.

\begin{table*}[htbp]
\renewcommand\arraystretch{1}
  \centering
  \begin{tabular}{|p{4cm}<{\centering}|p{4cm}<{\centering}|p{4cm}<{\centering}|}\hline
  $b|g_m$  & Alice's Measurements  \\\hline
  $0|g_1$ & $(\sqrt{2}/\sqrt{3},\,0,\,1/\sqrt{3})$  \\\hline
  $0|g_2$ & $(0,\,-\sqrt{2}/\sqrt{3},\,1/\sqrt{3})$  \\\hline
  $1|g_1$ & $(-\sqrt{2}/\sqrt{3},\,0,\,1/\sqrt{3})$  \\\hline
  $1|g_2$ & $(0,\,\sqrt{2}/\sqrt{3},\,1/\sqrt{3})$  \\\hline
\end{tabular}
  \caption{For Werner states, Alice's measurement directions are chosen based on $b|g_m$ in the four-setting case.}\label{tstra4}
\end{table*}

$\bullet$ For $n=6$ (see Fig. \ref{610setting} (a)),
\begin{equation}\label{logic}
\begin{split}
&E=\left(
\begin{array}{cc}
\frac{1}{2}\sqrt{2+\sqrt{2-\frac{2}{\sqrt{5}}}} & -\frac{1}{2}\sqrt{2-\sqrt{2-\frac{2}{\sqrt{5}}}}\\
\frac{1}{2}\sqrt{2-\sqrt{2-\frac{2}{\sqrt{5}}}} & \frac{1}{2}\sqrt{2+\sqrt{2-\frac{2}{\sqrt{5}}}}\\
\end{array}
\right),\\
&g_1=\left(
\begin{array}{cc}
1 & 0\\
0 & 1\\
\end{array}
\right),
g_2=\left(
\begin{array}{cc}
\frac{1}{\sqrt{2}} & \frac{-i}{\sqrt{2}}\\
\frac{1}{\sqrt{2}} & \frac{i}{\sqrt{2}}\\
\end{array}
\right),
g_3=\left(
\begin{array}{cc}
\frac{1}{\sqrt{2}} & \frac{1}{\sqrt{2}}\\
\frac{i}{\sqrt{2}} & \frac{-i}{\sqrt{2}}\\
\end{array}
\right).
\end{split}
\end{equation}

\begin{table*}[htbp]
\renewcommand\arraystretch{1}
  \centering
  \begin{tabular}{|p{4cm}<{\centering}|p{4cm}<{\centering}|p{5cm}<{\centering}|}\hline
  Guessing Strategy  & Optimal Input State & Success Probability \\\hline
  $(b,\,b,\,0)$ & $(\alpha,\,0,\,\beta)$ & $(\frac{15+\sqrt{5}}{20},\,\frac{5+\sqrt{5}}{10},\,\frac{5+\sqrt{5}}{10})$ \\\hline
  $(b\oplus 1,\,b\oplus 1,\,1)$ & $(-\alpha,\,0,\,-\beta)$ & $(\frac{15+\sqrt{5}}{20},\,\frac{5+\sqrt{5}}{10},\,\frac{5+\sqrt{5}}{10})$\\\hline
  $(b\oplus 1,\,b,\,1)$ & $(-\alpha,\,0,\,\beta)$ & $(\frac{15+\sqrt{5}}{20},\,\frac{5+\sqrt{5}}{10},\,\frac{5+\sqrt{5}}{10})$\\\hline
  $(b,\,b\oplus 1,\,0)$ & $(\alpha,\,0,\,-\beta)$ & $(\frac{15+\sqrt{5}}{20},\,\frac{5+\sqrt{5}}{10},\,\frac{5+\sqrt{5}}{10})$\\\hline
  $(b,\,0,\,b)$ & $(\beta,\,\alpha,\,0)$ & $(\frac{5+\sqrt{5}}{10},\,\frac{5+\sqrt{5}}{10},\,\frac{15+\sqrt{5}}{20})$\\\hline
  $(b\oplus 1,\,1,\,b\oplus 1)$ & $(-\beta,\,-\alpha,\,0)$ & $(\frac{5+\sqrt{5}}{10},\,\frac{5+\sqrt{5}}{10},\,\frac{15+\sqrt{5}}{20})$\\\hline
  $(b\oplus 1,\,0,\,b)$ & $(-\beta,\,\alpha,\,0)$ & $(\frac{5+\sqrt{5}}{10},\,\frac{5+\sqrt{5}}{10},\,\frac{15+\sqrt{5}}{20})$\\\hline
  $(b,\,1,\,b\oplus 1)$ & $(\beta,\,-\alpha,\,0)$ & $(\frac{5+\sqrt{5}}{10},\,\frac{5+\sqrt{5}}{10},\,\frac{15+\sqrt{5}}{20})$\\\hline
  $(0,\,b,\,b)$ & $(0,\,\beta,\,\alpha)$ & $(\frac{5+\sqrt{5}}{10},\,\frac{15+\sqrt{5}}{20},\,\frac{5+\sqrt{5}}{10})$\\\hline
  $(1,\,b\oplus 1,\,b\oplus 1)$ & $(0,\,-\beta,\,-\alpha)$ & $(\frac{5+\sqrt{5}}{10},\,\frac{15+\sqrt{5}}{20},\,\frac{5+\sqrt{5}}{10})$\\\hline
  $(0,\,b,\,b\oplus 1)$ & $(0,\,-\beta,\,\alpha)$ & $(\frac{5+\sqrt{5}}{10},\,\frac{15+\sqrt{5}}{20},\,\frac{5+\sqrt{5}}{10})$\\\hline
  $(1,\,b\oplus 1,\,b)$ & $(0,\,\beta,\,-\alpha)$ & $(\frac{5+\sqrt{5}}{10},\,\frac{15+\sqrt{5}}{20},\,\frac{5+\sqrt{5}}{10})$\\\hline
\end{tabular}
  \caption{The guessing strategy using the single-qubit state in the six-setting case. $\alpha=\sqrt{50+10\sqrt{5}}/10$, $\beta=\sqrt{50-10\sqrt{5}}/10$. In the column of success probability, three values correspond to different $m$. For example, in the first row, when $m=1$, the maximum probability is $(15+\sqrt{5})/20$; when $m=2$, the probability is $(5+\sqrt{5})/10$; when $m=3$, the probability is $(5+\sqrt{5})/10$. The corresponding success probability using the optimal input state is $P^s=\frac{1}{3}(\frac{15+\sqrt{5}}{20}+\frac{5+\sqrt{5}}{10}+\frac{5+\sqrt{5}}{10})=\frac{7+\sqrt{5}}{12}$. The other expressions in this table are similar with Table \ref{stra4}.}\label{stra6}
\end{table*}

The single-qubit strategies are shown in Table \ref{stra6}. Taking the first row for example, the success probability is calculated as follows,.
When $m=1$ and guessing $j=b$, the success probability is
\begin{equation}
\text{Tr}[U(|0\rangle\langle0|\otimes \,\rho)U^\dag(\Pi_0^Z\otimes\Pi_0^Z+\Pi_1^Z\otimes\Pi_1^Z)]=\frac{15+\sqrt{5}}{20}.
\end{equation}
When $m=2$ and guessing $j=b$, the success probability is
\begin{equation}
\text{Tr}[U(|0\rangle\langle0|\otimes g_2\, \cdot\rho\cdot\, g_2^\dag)U^\dag(\Pi_0^Z\otimes\Pi_0^Z+\Pi_1^Z\otimes\Pi_1^Z)]=\frac{5+\sqrt{5}}{10}.
\end{equation}
When $m=3$ and guessing $j=0$, the success probability is
\begin{equation}
\text{Tr}[U(|0\rangle\langle0|\otimes g_3 \,\cdot\rho\cdot\, g_3^\dag)U^\dag (\Pi_0^Z\otimes \mathbb{I})]=\frac{5+\sqrt{5}}{10}.
\end{equation}

The success probability using $\rho$ is $P^s_\rho=\frac{1}{3}(\frac{15+\sqrt{5}}{20}+\frac{5+\sqrt{5}}{10}+\frac{5+\sqrt{5}}{10})=\frac{7+\sqrt{5}}{12}$. As $\rho$ is the optimal input state, the bound of single-qubit protocol is $P^s_6=\frac{7+\sqrt{5}}{12}$. The other rows are similar.

The details of Alice's measurement settings when using the Werner state in the six-setting case are introduced in Table \ref{tstra6}.

\begin{table*}[htbp]
\renewcommand\arraystretch{1}
  \centering
  \begin{tabular}{|p{4cm}<{\centering}|p{4cm}<{\centering}|p{4cm}<{\centering}|}\hline
  $b|g_m$  & Alice's Measurements  \\\hline
  $0|g_1$ & $(\alpha,\,0,\,\beta)$  \\\hline
  $0|g_2$ & $(0,\,-\beta,\,\alpha)$  \\\hline
  $0|g_3$ & $(\beta,\,-\alpha,\,0)$  \\\hline
  $1|g_1$ & $(-\alpha,\,0,\,\beta)$  \\\hline
  $1|g_2$ & $(0,\,-\beta,\,-\alpha)$  \\\hline
  $1|g_3$ & $(\beta,\,\alpha,\,0)$  \\\hline
\end{tabular}
  \caption{For Werner states, Alice's measurement directions are chosen based on $b|g_m$ in the six-setting case.}\label{tstra6}
\end{table*}

\begin{table*}[htbp]
\renewcommand\arraystretch{1.25}
  \centering
  \begin{tabular}{|p{4.5cm}<{\centering}|p{5cm}<{\centering}|p{4.5cm}<{\centering}|}\hline
  Guessing Strategy  & Optimal Input State & Success Probability \\\hline
  $(b,\,b,\,b\oplus 1,\,b\oplus 1,\,0)$ & $(\frac{\sqrt{2}}{\sqrt{3}},\,0,\,\frac{1}{\sqrt{3}})$ & $(\frac{5}{6},\,\frac{2}{3},\,\frac{7+\sqrt{5}}{12},\,\frac{3+\sqrt{5}}{6},\,\frac{2}{3})$ \\\hline
  $(b,\,b\oplus 1,\,b\oplus 1,\,0,\,b)$ & $(\frac{1}{\sqrt{6}},\,\frac{\sqrt{5}}{\sqrt{6}},\,0)$ & $(\frac{2}{3},\,\frac{7+\sqrt{5}}{12},\,\frac{3+\sqrt{5}}{6},\,\frac{2}{3},\,\frac{5}{6})$ \\\hline
  $(b\oplus 1,\,b\oplus 1,\,0,\,b,\,b)$ & $(-\frac{\sqrt{2}}{\sqrt{15}-\sqrt{3}},\,\frac{\sqrt{2}}{\sqrt{15}-\sqrt{3}},\,-\frac{\sqrt{5}-1}{2\sqrt{3}})$ & $(\frac{7+\sqrt{5}}{12},\,\frac{3+\sqrt{5}}{6},\,\frac{2}{3},\,\frac{5}{6},\,\frac{2}{3})$ \\\hline
  $(b\oplus 1,\,0,\,b,\,b,\,b\oplus 1)$ & $(-\frac{\sqrt{5}}{\sqrt{6}},\,-\frac{1}{\sqrt{6}},\,0)$ & $(\frac{3+\sqrt{5}}{6},\,\frac{2}{3},\,\frac{5}{6},\,\frac{2}{3},\,\frac{7+\sqrt{5}}{12})$ \\\hline
  $(0,\,b,\,b,\,b\oplus 1,\,b\oplus 1)$ & $(0,\,-\frac{\sqrt{2}}{\sqrt{3}},\,\frac{1}{\sqrt{3}})$ & $(\frac{2}{3},\,\frac{5}{6},\,\frac{2}{3},\,\frac{7+\sqrt{5}}{12},\,\frac{3+\sqrt{5}}{6})$ \\\hline
  $(b\oplus 1,\,b\oplus 1,\,b,\,b,\,1)$ & $(-\frac{\sqrt{2}}{\sqrt{3}},\,0,\,-\frac{1}{\sqrt{3}})$ & $(\frac{5}{6},\,\frac{2}{3},\,\frac{7+\sqrt{5}}{12},\,\frac{3+\sqrt{5}}{6},\,\frac{2}{3})$ \\\hline
  $(b\oplus 1,\,b,\,b,\,1,\,b\oplus 1)$ & $(-\frac{1}{\sqrt{6}},\,-\frac{\sqrt{5}}{\sqrt{6}},\,0)$ & $(\frac{2}{3},\,\frac{7+\sqrt{5}}{12},\,\frac{3+\sqrt{5}}{6},\,\frac{2}{3},\,\frac{5}{6})$ \\\hline
  $(b,\,b,\,1,\,b\oplus 1,\,b\oplus 1)$ & $(\frac{\sqrt{2}}{\sqrt{15}-\sqrt{3}},\,-\frac{\sqrt{2}}{\sqrt{15}-\sqrt{3}},\,\frac{\sqrt{5}-1}{2\sqrt{3}})$ & $(\frac{7+\sqrt{5}}{12},\,\frac{3+\sqrt{5}}{6},\,\frac{2}{3},\,\frac{5}{6},\,\frac{2}{3})$ \\\hline
  $(b,\,1,\,b\oplus 1,\,b\oplus 1,\,b)$ & $(\frac{\sqrt{5}}{\sqrt{6}},\,\frac{1}{\sqrt{6}},\,0)$ & $(\frac{3+\sqrt{5}}{6},\,\frac{2}{3},\,\frac{5}{6},\,\frac{2}{3},\,\frac{7+\sqrt{5}}{12})$ \\\hline
  $(1,\,b\oplus 1,\,b\oplus 1,\,b,\,b)$ & $(0,\,\frac{\sqrt{2}}{\sqrt{3}},\,-\frac{1}{\sqrt{3}})$ & $(\frac{2}{3},\,\frac{5}{6},\,\frac{2}{3},\,\frac{7+\sqrt{5}}{12},\,\frac{3+\sqrt{5}}{6})$ \\\hline
  $(b\oplus 1,\,0,\,0,\,0,\,b\oplus 1)$ & $(-\frac{\sqrt{2}}{\sqrt{3}},\,0,\,\frac{1}{\sqrt{3}})$ & $(\frac{5}{6},\,\frac{7+\sqrt{5}}{12},\,\frac{7+\sqrt{5}}{12},\,\frac{2}{3},\,\frac{7+\sqrt{5}}{12})$ \\\hline
  $(0,\,0,\,0,\,b\oplus 1,\,b\oplus 1)$ & $(-\frac{\sqrt{5}-1}{2\sqrt{6}},\,-\frac{\sqrt{5}-1}{2\sqrt{6}},\,\frac{2}{\sqrt{15}-\sqrt{3}})$ & $(\frac{7+\sqrt{5}}{12},\,\frac{7+\sqrt{5}}{12},\,\frac{2}{3},\,\frac{7+\sqrt{5}}{12},\,\frac{5}{6})$ \\\hline
  $(0,\,0,\,b\oplus 1,\,b\oplus 1,\,0)$ & $(\frac{\sqrt{5}-1}{2\sqrt{6}},\,\frac{\sqrt{5}-1}{2\sqrt{6}},\,\frac{2}{\sqrt{15}-\sqrt{3}})$ & $(\frac{7+\sqrt{5}}{12},\,\frac{2}{3},\,\frac{7+\sqrt{5}}{12},\,\frac{5}{6},\,\frac{7+\sqrt{5}}{12})$ \\\hline
  $(0,\,b\oplus 1,\,b\oplus 1,\,0,\,0)$ & $(0,\,\frac{\sqrt{2}}{\sqrt{3}},\,\frac{1}{\sqrt{3}})$ & $(\frac{2}{3},\,\frac{7+\sqrt{5}}{12},\,\frac{5}{6},\,\frac{7+\sqrt{5}}{12},\,\frac{7+\sqrt{5}}{12})$ \\\hline
  $(b\oplus 1,\,b\oplus 1,\,0,\,0,\,0)$ & $(-\frac{\sqrt{2}}{\sqrt{15}-\sqrt{3}},\,\frac{\sqrt{2}}{\sqrt{15}-\sqrt{3}},\,\frac{\sqrt{5}-1}{2\sqrt{3}})$ & $(\frac{7+\sqrt{5}}{12},\,\frac{5}{6},\,\frac{7+\sqrt{5}}{12},\,\frac{7+\sqrt{5}}{12},\,\frac{2}{3})$ \\\hline
  $(b,\,1,\,1,\,1,\,b)$ & $(\frac{\sqrt{2}}{\sqrt{3}},\,0,\,-\frac{1}{\sqrt{3}})$ & $(\frac{5}{6},\,\frac{7+\sqrt{5}}{12},\,\frac{7+\sqrt{5}}{12},\,\frac{2}{3},\,\frac{7+\sqrt{5}}{12})$ \\\hline
  $(1,\,1,\,1,\,b,\,b)$ & $(\frac{\sqrt{5}-1}{2\sqrt{6}},\,\frac{\sqrt{5}-1}{2\sqrt{6}},\,-\frac{2}{\sqrt{15}-\sqrt{3}})$ & $(\frac{7+\sqrt{5}}{12},\,\frac{7+\sqrt{5}}{12},\,\frac{2}{3},\,\frac{7+\sqrt{5}}{12},\,\frac{5}{6})$ \\\hline
  $(1,\,1,\,b,\,b,\,1)$ & $(-\frac{\sqrt{5}-1}{2\sqrt{6}},\,-\frac{\sqrt{5}-1}{2\sqrt{6}},\,-\frac{2}{\sqrt{15}-\sqrt{3}})$ & $(\frac{7+\sqrt{5}}{12},\,\frac{2}{3},\,\frac{7+\sqrt{5}}{12},\,\frac{5}{6},\,\frac{7+\sqrt{5}}{12})$ \\\hline
  $(1,\,b,\,b,\,1,\,1)$ & $(0,\,-\frac{\sqrt{2}}{\sqrt{3}},\,-\frac{1}{\sqrt{3}})$ & $(\frac{2}{3},\,\frac{7+\sqrt{5}}{12},\,\frac{5}{6},\,\frac{7+\sqrt{5}}{12},\,\frac{7+\sqrt{5}}{12})$ \\\hline
  $(b,\,b,\,1,\,1,\,1)$ & $(\frac{\sqrt{2}}{\sqrt{15}-\sqrt{3}},\,-\frac{\sqrt{2}}{\sqrt{15}-\sqrt{3}},\,-\frac{\sqrt{5}-1}{2\sqrt{3}})$ & $(\frac{7+\sqrt{5}}{12},\,\frac{5}{6},\,\frac{7+\sqrt{5}}{12},\,\frac{7+\sqrt{5}}{12},\,\frac{2}{3})$ \\\hline
\end{tabular}
  \caption{The guessing strategy using the single-qubit state in the ten-setting case. The expressions in this table are the same with Table \ref{stra6}. The average success probability is $(13+\sqrt{5})/20$.}\label{stra10}
\end{table*}

$\bullet$ For $n=10$ (see Fig. \ref{610setting} (b)),
\begin{equation}\label{logic}
\begin{split}
&E=\left(
\begin{array}{cc}
\sqrt{\frac{3+\sqrt{3}}{6}} & -\sqrt{\frac{3-\sqrt{3}}{6}}\\
\sqrt{\frac{3-\sqrt{3}}{6}} & \sqrt{\frac{3+\sqrt{3}}{6}}\\
\end{array}
\right),\,
g_1=\left(
\begin{array}{cc}
1 & 0\\
0 & 1\\
\end{array}
\right),\\
&g_2=\left(
\begin{array}{cc}
\frac{1+\sqrt{5}-2 i}{4} & -\frac{(1-i)(\sqrt{5}-1)}{4\sqrt{2}}\\
\frac{(1+i)(\sqrt{5}-1)}{4\sqrt{2}} & \frac{1+\sqrt{5}+2 i}{4}\\
\end{array}
\right),\\
&g_3=g_2\cdot g_2,\,
g_4=g_2\cdot g_2\cdot g_2,\,
g_5=g_2\cdot g_2\cdot g_2\cdot g_2.
\end{split}
\end{equation}

The single-qubit strategies are shown in Table \ref{stra10}. Here we show the success probability of each $m$, and the upper-bound success probability is $P^s_{10}=(13+\sqrt{5})/20$. Since $\eta^*_{10}=(3+\sqrt{5})/10\approx0.524$, we have $P^s_{10}=(1+\eta^*_{10})/2$.

\begin{table*}[htbp]
\renewcommand\arraystretch{1}
  \centering
  \begin{tabular}{|p{4cm}<{\centering}|p{6cm}<{\centering}|p{4cm}<{\centering}|}\hline
  $b|g_m$  & Alice's Measurements  \\\hline
  $0|g_1$ & $(\frac{\sqrt{2}}{\sqrt{3}},\,0,\,\frac{1}{\sqrt{3}})$  \\\hline
  $0|g_2$ & $(0,\,\frac{\sqrt{2}}{\sqrt{3}},\,\frac{1}{\sqrt{3}})$  \\\hline
  $0|g_3$ & $(-\frac{\sqrt{5}}{\sqrt{6}},\,\frac{1}{\sqrt{6}},\,0)$  \\\hline
  $0|g_4$ & $(-\frac{\sqrt{2}}{\sqrt{15}-\sqrt{3}},\,-\frac{\sqrt{2}}{\sqrt{15}-\sqrt{3}},\,-\frac{\sqrt{5}-1}{2\sqrt{3}})$  \\\hline
  $0|g_5$ & $(\frac{1}{\sqrt{6}},\,-\frac{\sqrt{5}}{\sqrt{6}},\,0)$  \\\hline
  $1|g_1$ & $(-\frac{\sqrt{2}}{\sqrt{3}},\,0,\,\frac{1}{\sqrt{3}})$  \\\hline
  $1|g_2$ & $(-\frac{\sqrt{2}}{\sqrt{15}-\sqrt{3}},\,-\frac{\sqrt{2}}{\sqrt{15}-\sqrt{3}},\,\frac{\sqrt{5}-1}{2\sqrt{3}})$  \\\hline
  $1|g_3$ & $(0,\,-\frac{\sqrt{2}}{\sqrt{3}},\,\frac{1}{\sqrt{3}})$  \\\hline
  $1|g_4$ & $(\frac{\sqrt{5}-1}{2\sqrt{6}},\,-\frac{\sqrt{5}-1}{2\sqrt{6}},\,\frac{2}{\sqrt{15}-\sqrt{3}})$  \\\hline
  $1|g_5$ & $(-\frac{\sqrt{5}-1}{2\sqrt{6}},\,\frac{\sqrt{5}-1}{2\sqrt{6}},\,\frac{2}{\sqrt{15}-\sqrt{3}})$  \\\hline
\end{tabular}
  \caption{For Werner states, Alice's measurement directions are chosen based on $b|g_m$ in the ten-setting case.}\label{tstra10}
\end{table*}

The details of Alice's measurement settings when using the Werner state in the ten-setting case are introduced in Table \ref{tstra10}.

In summary, for the multi-setting cases, $m$ unitary gates are employed to realize $n$ measurement settings. When $n=2,\,4,\,6,\,10$, one measurement setting corresponds to one value of $b|g_m$ and the number of gates $m=n/2$ as introduced above. When $n=3$, the situation is different and we still need $m=3$ unitary gates before the operation $K_{ij}$ for the convenience in both theory and experiment. There are six values of $b|g_m$ in the three-setting case, while two values of them correspond to the same measurement setting. As shown in Fig. \ref{34setting} (a), there exist three lines crossing through three antipodal pairs of vertices of the octahedron, i.e., two along the diagonal and anti-diagonal directions in the $x-z$ plane and one in the direction along the $y$ axis. As listed in Table \ref{tstra3}, the diagonal direction in the $x-z$ plane corresponds to the results $0|g_1$ and $0|g_2$ , the anti-diagonal direction in the $x-z$ plane corresponds to the results $1|g_1$ and $1|g_3$, the $y$ direction corresponds to the results $1|g_2$ and $0|g_3$.

\subsection{The SD method for the Bell diagonal states}

\begin{figure}[!h]
\centering
\includegraphics[width=0.4\textwidth]{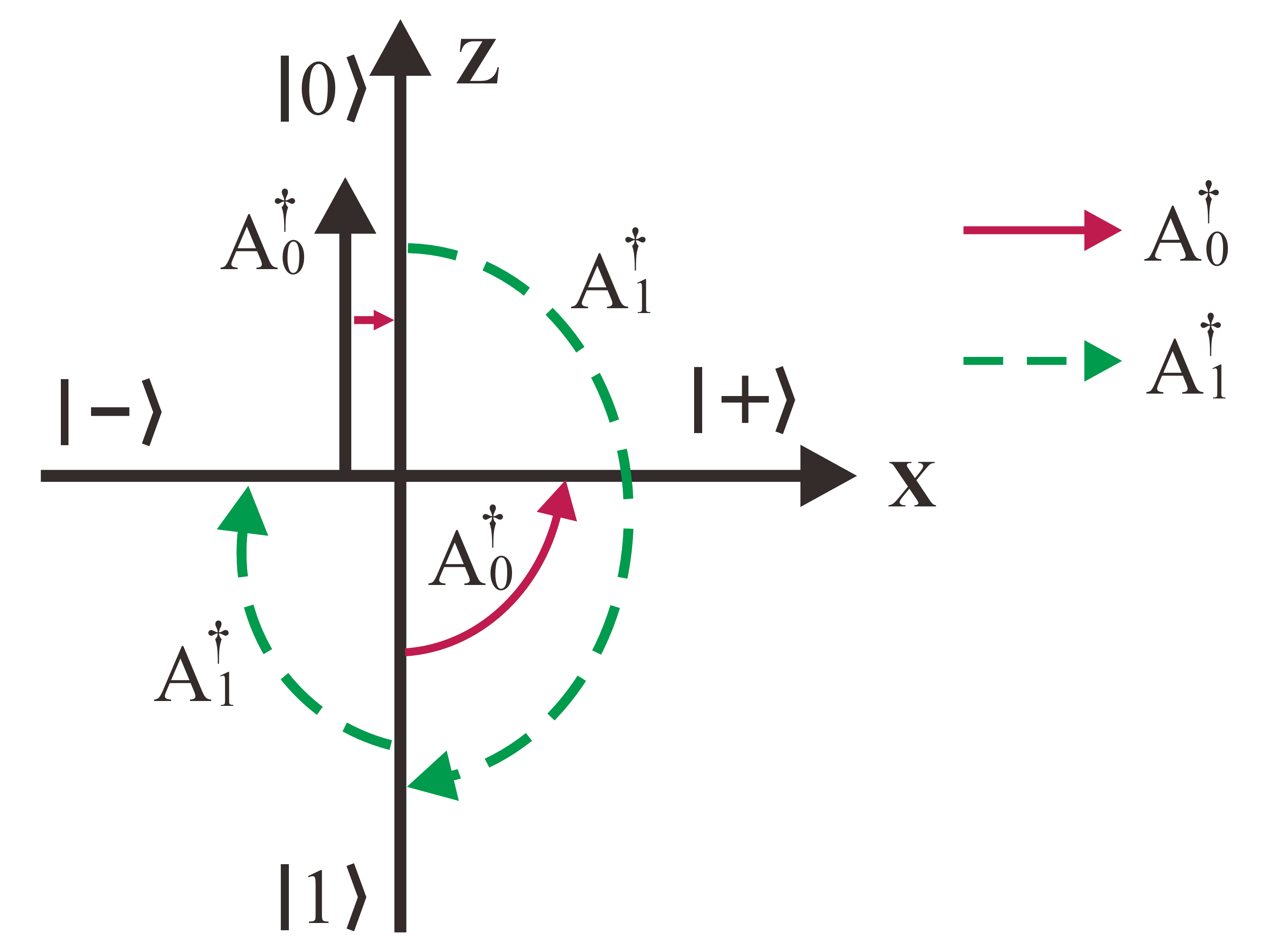}
\caption{Illustrating the functions of $A_0^\dag$ and $A_1^\dag$ for the Bell diagonal states.
}\label{ffigs1}
\end{figure}

\begin{figure*}[htbp]
\centering
\includegraphics[width=0.7\textwidth]{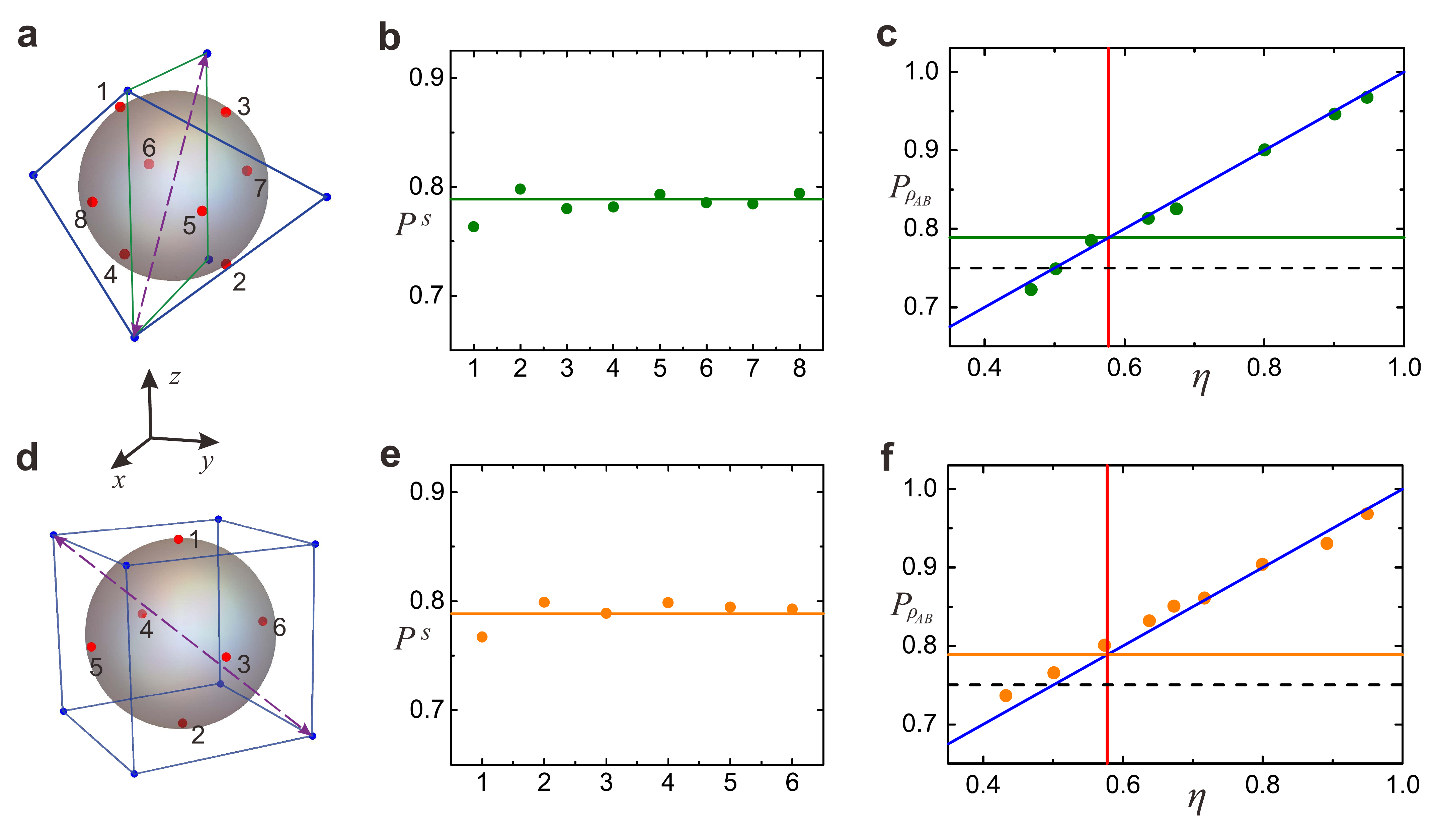}
\caption{The results for three and four measurement settings. (a) and (d) One measurement setting is shown as a purple arrowed lines in the Bloch spheres for each case (three and four measurement settings, respectively). The blue points represent the vertices of the corresponding Platonic solids. The red points located at the center of every face represent the initial single-qubit states which are used to obtain the upper bounds for the single-qubit protocol. (b) and (e) The results for the single-qubit upper bounds for three and four measurement settings, respectively, with the corresponding initial single-qubit states. The points represent the experimental results for the states labeled on the corresponding Platonic solids, whereas the lines represent the theoretical predictions, which are identical for the cases of three and four measurement settings. (c) and (f) The results for two-qubit states. The green and orange points represent the experimental results obtained with three and four measurement settings, respectively. The green and orange solid lines represent the upper bounds of the single-qubit protocol for three and four measurement settings, respectively. The black dashed lines represent the theoretical results for an infinite number of measurement settings.}\label{34setting}
\end{figure*}

\begin{figure*}[htbp]
\centering
\includegraphics[width=0.6\textwidth]{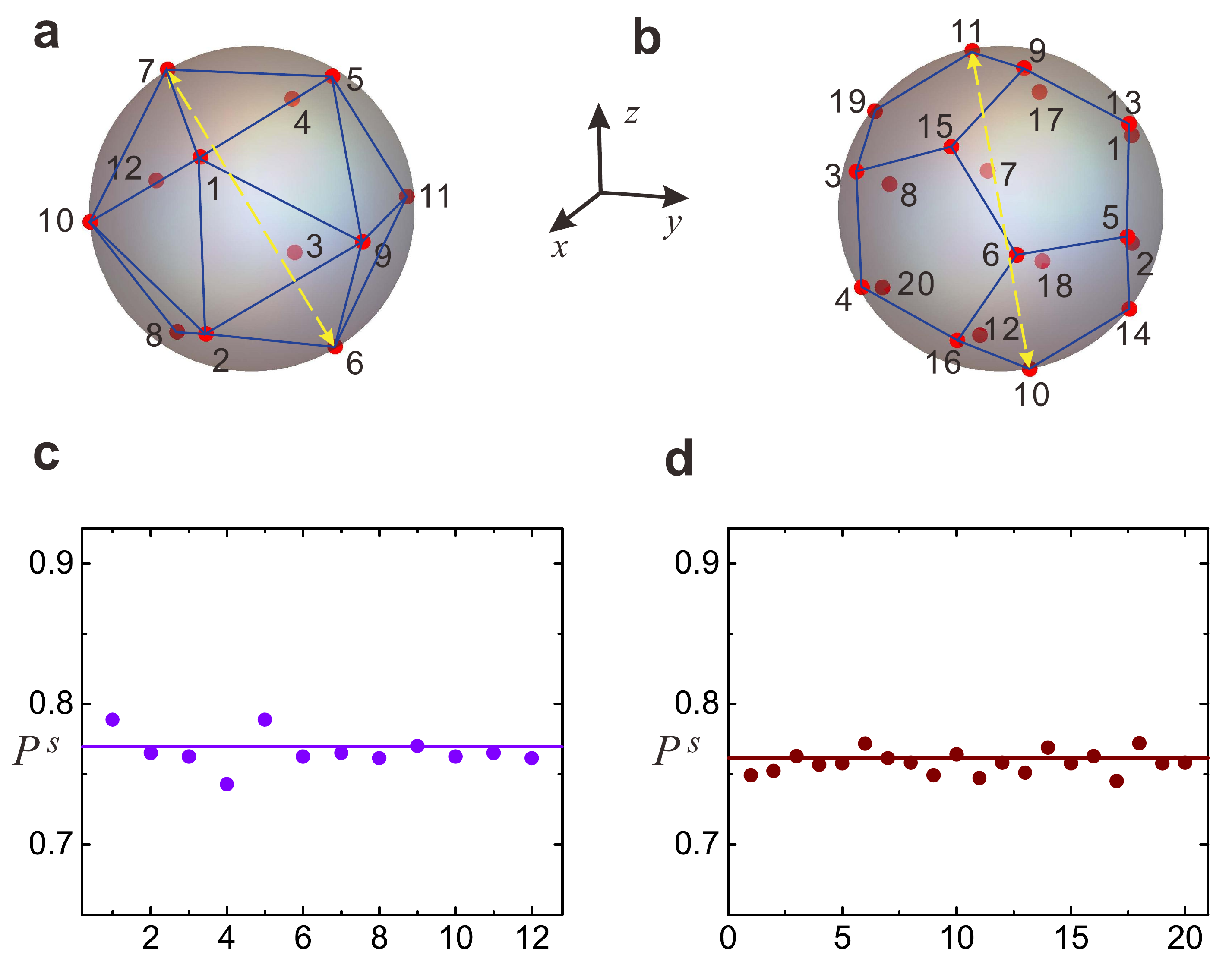}
\caption{The results for six and ten measurement settings. (a) and (b) One measurement setting is shown as a yellow arrowed lines in the Bloch spheres for each case (six and ten measurement settings, respectively). The red points represent the vertices of the corresponding Platonic solids and also represent the initial single-qubit states used to obtain the single-qubit upper bounds. (c) and (d) The results for single-qubit upper bounds for six and ten measurement settings, respectively, for the states labeled on the corresponding Platonic solids. The purple and dark red points represent the experimental results obtained with six and ten measurement settings, respectively, and the purple and dark red solid line represent the corresponding theoretical prediction.}\label{610setting}
\end{figure*}

To present a well-understood of how the SD method works, let us take the Bell diagonal states for example. The Bell diagonal states can be expressed as $\rho_{BD}=\frac{1}{4}(\mathbb{I} + \sum_{i=x,y,z} t_i \; \sigma_i \otimes \sigma_i)$, where $\mathbb{I}$ is a $4\times 4$ unit matrix, $\sigma_i$ is the Pauli matrix and $(t_x,t_y,t_z)$ is a real parameter vector which belongs to the tetrahedron defined by the set of vertices $(-1,-1,-1),(-1,1,1),(1,-1,1)$ and $(1,1,-1)$. For the sake of convenience in the demonstration, let's take $t_x = t_y$, then the vector of $(t_x,t_z)$ is inside the triangle with vertices $(-1,-1),(0,1),(1,-1)$. The steering ellipsoid of $\rho_{BD}$ after Alice's measurement is $\frac{x^2}{t_x^2}+\frac{y^2}{t_x^2}+\frac{z^2}{t_z^2}=1$. Similar with the method used in the \textbf{Section A}, $A_j^\dag$ should satisfy: $A_0^\dag$ turns $|0\rangle$ to $|0\rangle$ and turns $|1\rangle$ to $|+\rangle|$, $A_1^\dag$ turns $|0\rangle$ to $|1\rangle$ and turns $|1\rangle$ to $|-\rangle$, as shown in Fig. \ref{ffigs1}. We have,

\begin{equation}\label{lhs}
\left\{
\begin{aligned}
& A_0^\dag \cdot |0\rangle\langle0| \cdot A_0 \varpropto |0\rangle\langle0| \\
& A_1^\dag \cdot |0\rangle\langle0| \cdot A_1 \varpropto |1\rangle\langle1| \\
& A_0^\dag \cdot |1\rangle\langle1| \cdot A_0 \varpropto |+\rangle\langle+| \\
& A_1^\dag \cdot |1\rangle\langle1| \cdot A_1 \varpropto |-\rangle\langle-| \\
\end{aligned}
\right.
\end{equation}
which gives,
\begin{equation}\label{ssubc2}
A_{0}=\left(\begin{array}{cc}
\frac{1}{\sqrt{2}} & 0\\
\frac{1}{2} & \frac{1}{2}\\
\end{array}\right),\,\,\,\,
A_{1}=\left(\begin{array}{cc}
0 & \frac{1}{\sqrt{2}}\\
\frac{1}{2} & -\frac{1}{2}\\
\end{array}\right).
\end{equation}

The corresponding subchannels $K_{ij}$ are
\begin{equation}\label{ssubc1}
\begin{split}
K_{00}=\left(\begin{array}{cc}
\frac{1}{\sqrt{2}} & 0\\
0 & 0\\
\end{array}\right),\,\,\,
K_{01}=\left(\begin{array}{cc}
0 & \frac{1}{\sqrt{2}}\\
0 & 0\\
\end{array}\right),\\
K_{10}=\left(\begin{array}{cc}
0 & 0\\
\frac{1}{2} & \frac{1}{2}\\
\end{array}\right),\,\,\,\,
K_{11}=\left(\begin{array}{cc}
0 & 0\\
\frac{1}{2} & -\frac{1}{2}\\
\end{array}\right).
\end{split}
\end{equation}

By directly calculating, we get the maximum probability of the single-qubit is $P^s=(1+1/\sqrt{2})/2$. One of the optimal strategy for single-qubit is input $\rho=\frac{1}{2}(\mathbb{I} + \vec{n} \cdot \vec{\sigma})$ where $\vec{n}=(\frac{1}{\sqrt{2}},0,\frac{1}{\sqrt{2}})$ and the strategy is always guess $j=0$. For the Bell diagonal states, the guessing strategy is based on the result $b$, if $b=0$, then measuring along $(0,\ 0,\ -1)$ direction, if $b=1$, then measuring along $(1,\ 0,\ 0)$ direction. The measurement result is $a$, then guess $j=a$. The successful probability is $P_{\rho_{BD}}= \frac{1}{4}(2+t_x-t_z)$. Thus, for the states satisfy $t_x-t_z>\sqrt{2}$, the steering of Alice to Bob is certified by this SD task.

\section{Experimental implementation of $\{g_m\}$}

To realize $\{g_m\}$ in the multi-setting cases, several wave plates including HWPs and quarter-wave plates (QWPs) are employed which can be written in Jones matrix form as below,
\begin{equation}\label{hqwp}
\begin{split}
  &J_h=\left(
\begin{array}{cc}
\cos 2\phi_h & -\sin 2\phi_h\\
\sin 2\phi_h & \cos 2\phi_h\\
\end{array}
\right),\\
&J_q=\left(
\begin{array}{cc}
\cos^2\phi_q + i \sin^2\phi_q & 0.5 (1-i)\sin 2\phi_q\\
0.5 (1-i)\sin 2\phi_q & i \cos^2\phi_q + \sin^2\phi_q\\
\end{array}
\right),
\end{split}
\end{equation} where $\phi_h$ and $\phi_q$ are the angle settings for the HWP and QWP, respectively.

\begin{itemize}
\item For $n=3$, the gates $g_2$ and $g_3$ are realized by using a QWP with $\phi_q=-3\pi/8$ and $-\pi/8$, respectively.
\item For $n=4$, $g_2$ is realized using a QWP with $\phi_q=\pi/2$.
\item For $n=6$, an HWP followed by a QWP is employed to realize gates $g_2$ and $g_3$. For $g_2$, $\phi_h=\pi/8$ and $\phi_q=0$; for $g_3$, $\phi_h=\pi/8$ and $\phi_q=-\pi/4$.
\item For $n=10$, a combination consisting of two QWPs and an HWP can be used to implement the gates $g_i$ ($i=2,\,3,\,4,\,5$). QWP1, HWP and QWP2 are placed in sequence, and the details of the degree settings $\phi_{q1},\,\phi_h$, and $\phi_{q2}$ are as listed in the table \ref{ten}.
\begin{table*}[htbp]
\renewcommand\arraystretch{1.5}
  \centering
  \begin{tabular}{|p{2cm}<{\centering}|p{2cm}<{\centering}|p{2cm}<{\centering}|p{2cm}<{\centering}|}\hline
gate & $\phi_{q1}$ & $\phi_h$ & $\phi_{q2}$ \\ \hline
$g_2$ & $25.6^\circ$ & $49.7^\circ$ & $40.8^\circ$ \\ \hline
$g_3$ & $8.8^\circ$ & $64.2^\circ$ & $57.6^\circ$ \\ \hline
$g_4$ & $-32.4^\circ$ & $64.2^\circ$ & $-81.2^\circ$ \\ \hline
$g_5$ & $-49.2^\circ$ & $49.7^\circ$ & $-64.4^\circ$ \\ \hline
\end{tabular}
  \caption{Realization of the gates with the degrees of $\phi_{q1},\,\phi_h,\,\phi_{q2}$.}\label{ten}
\end{table*}
\end{itemize}

\section{More experimental results}

In this work, the fidelity of the experimental state $\rho_e$ and the target entangled state $\rho_t$ is $F=[Tr[\sqrt{\sqrt{\rho_t}\cdot\rho_e \cdot\sqrt{\rho_t}}]]^2$. By performing the state tomography, we obtain $\rho_e$. When fitting $\rho_e$ with Werner states which means $\rho_t=\rho_{AB}$ in Eq. (2) in the main text, we maximize $F$ to determine the value of $\eta$. For the maximally entangled state we prepared, the fidelity 95.5\% is obtained by choosing $\rho_t=|\Phi\rangle\langle\Phi|$. When fitting this experimental state with Werner state, we get $\eta=0.947$ which should be $1$ in theory and the corresponding fidelity is about 97.4\%. Since we prepared several experimental Werner states, for every state, we obtained the corresponding $\eta$ and fidelity. The value 98\% is the average fidelity.

More experimental results of multi-setting cases are presented here. The detailed measurement settings are illustrated in the corresponding figures. In each case, the initial single-qubit states that are used to obtain the upper bounds are pure states located on the surface of the Bloch sphere. For the cases of three and four measurement settings, they are the centers of the faces of the corresponding octahedron and cube, respectively, whereas for the cases of six and ten measurement settings, they are the vertices of the corresponding icosahedron and dodecahedron, respectively.

\end{document}